\documentclass[11pt,a4paper]{article}

\usepackage{jheppub}
\usepackage{amsmath}
\usepackage{amssymb,amsthm}
\usepackage{mathtools}
\usepackage{mathrsfs}
\usepackage{relsize}
\usepackage{bm}
\usepackage{lipsum}
\usepackage{ulem}

\usepackage{epsfig,graphics,graphicx,color,xcolor}
\usepackage{soul} 
\usepackage{cancel} 
\usepackage{slashed}	
\usepackage{subfigure}
\usepackage{array}
\usepackage{ragged2e}
\usepackage{lineno}
\usepackage{natbib}
\usepackage{hyperref}
\hypersetup{colorlinks=true,linkcolor=blue,anchorcolor=blue,citecolor=darkgreen, filecolor=blue,urlcolor=blue,bookmarksnumbered=true,
pdfview=FitB
}
\usepackage{enumitem}

\colorlet{darkgreen}{green!50!black}
\colorlet{brightyellow}{yellow!75!red}
\colorlet{orange}{red!50!yellow}
\colorlet{darkgray}{gray!50!black}

\def\dd{{\mathrm{d}}}

\makeatletter
\newcommand*{\transpose}{%
  {\mathpalette\@transpose{}}%
}
\newcommand*{\@transpose}[2]{%
  \raisebox{\depth}{$\m@th#1\intercal$}%
}
\makeatother 

\title{Covariant analysis of electromagnetic current on the light cone: 
exposition with scalar Yukawa theory}

\author[1]{Wenyu Zhang,}

\author[1, 2, 3, *]{Yang~Li,}

\author[3]{James P. Vary}

\affiliation[1]{Department of Modern Physics, University of Science and Technology of China, Hefei, Anhui 230026, China }
\affiliation[2]{Anhui Center for fundamental sciences (Theoretical Physics), University of Science and Technology of China, Hefei, Anhui 230026, China}
\affiliation[3]{Department of Physics and Astronomy, Iowa State University, Ames, IA 50011, United States}
\note[*]{Corresponding author}


\arxivnumber{}

\date{\today}

\abstract{
We present the first systematic investigation of the Lorentz covariance of the charge form factor for a strongly coupled scalar theory in (3+1)-dimensions. Our results are based on the first non-perturbative solution of the scalar Yukawa theory with a Fock sector expansion including up to thee-particles (one mock nucleon plus two mock pions or two mock nucleons plus one mock anti-nucleon). The light-front Hamiltonian is constructed and renormalized using a Fock sector dependent scheme. The derived eigenvalue equation is then solved non-perturbatively to obtain the wave functions, which are then used to compute the current matrix element. 

We perform a covariant analysis of the current matrix element taking into account possible violation of the Poincaré symmetry due to the Fock sector truncation. The physical form factor depends on two boost invariants $\zeta, \Delta^2_\perp$, instead of the single Lorentz invariant $Q^2$. Instead of adopting the conventional Drell-Yan frame $\zeta = 0$, we evaluate the form factor in general frames, and use the frame dependence to quantitatively gauge the loss of the Lorentz covariance. Our numerical result shows that as more Fock sectors are included, the frame dependence reduces dramatically. In particular, the anti-nucleon degree of freedom plays an important role in the reduction of the frame dependence, even though it only takes a small portion within the state vector. We also find that there is no zero-mode contribution to the current for the scalar Yukawa theory. }

\begin{document}
\maketitle

\section{Introduction} \label{sect:introduction}

The charge form factors $F(Q^2)$ play a key role in unraveling the structures of hadrons \cite{Gross:2022hyw, Gao:2021sml, Perdrisat:2006hj, Punjabi:2015bba, Pacetti:2014jai, Miller:2010nz, Hand:1963zz, Hofstadter:1956qs, Alexandrou:2012da}. On the one hand, it is one of the best measured observables in experiments. On the other hand, many puzzles remain on the theory side. While at large momentum transfer $Q^2 \gg \Lambda_\textsc{qcd}^2$ the asymptote of  $F(Q^2)$ is dictated by perturbative QCD (pQCD), the small-$Q^2$ behavior of $F(Q^2)$, which determines the shape of the hadron, falls into the realm of non-perturbative QCD \cite{Lepage:1980fj, Isgur:1984jm, Isgur:1988iw, Isgur:1989cy, Sterman:1997sx}. Despite continuing advancement in theory and phenomenology, notably in lattice gauge theory, our understanding of QCD in the strong coupling regime remains incomplete. 
In this regard, the light-front Hamiltonian approach is poised to tackle this problem by combining the increase of the computational power and the development of theoretical techniques \cite{Bakker:2013cea}. 

The light front approach, dating back to the late 60s, was initially developed to analyze high-energy scattering processes in QCD \cite{Fubini:1964boa, Bjorken:1968dy, Feynman:1969ej, Kogut:1972di, Altarelli:1977zs, Frankfurt:1977vc, Lepage:1980fj, Altarelli:1981ax, Frankfurt:1981mk, Chernyak:1983ej, Frankfurt:1988nt, Frankfurt:1991rk, Goeke:2001tz, Braun:2003rp, Ivanov:2004ax, Kovchegov:2012mbw, Cruz-Santiago:2015dla, Sterman:2016etx}. 
It was later realized that formulating quantum field theory on the light front provides a formal framework to describe physics in the non-perturbative regime \cite{Dirac:1949cp, Weinberg:1966jm, Susskind:1967rg, Kogut:1969xa, Bjorken:1970ah, Brodsky:1973kb, Coester:1992cg, Fuda:1992uh, Zhang:1994ti, Burkardt:1995ct, Brodsky:1997de, Carbonell:1998rj, Miller:2000kv, Heinzl:2000ht, Garsevanishvili:2007jp, Martinovic:2007nee, Frederico:2010zh, Brodsky:2014yha, Bakker:2014cua, Hiller:2016itl, Ji:2020ect, Gross:2022hyw}. 
For example, the celebrated Drell-Yan-West formula equates the pion form factor $F_\pi(Q^2)$ to the overlap of wave functions on the light front $x^+ = 0$ as \cite{Drell:1969km, West:1970av}, 
\begin{equation}\label{eqn:DYW}
F_\pi(Q^2) = \int_0^1\frac{\dd x}{2x(1-x)}\int\frac{\dd^2 k_\perp}{(2\pi)^3}
\psi^*_\pi(x, \vec k_\perp) \psi_\pi(x, \vec k_\perp-x \vec q_\perp) + \cdots
\end{equation}
where, $q^\mu$ is the 4-momentum of the probing photon, and $Q^2 = -q^2$. $\psi_\pi(x, \vec k_\perp)$ is the pion valence light-front wave function (LFWF), which only depends on boost invariants, viz. the longitudinal momentum fraction $x = x_1 = p_1^+/p^+$ of the quark and the relative transverse momentum $\vec k_\perp = \vec p_{1\perp} - x_1 \vec p_\perp$ of the quark. Here, we have adopted the light cone coordinates, $x^\pm = x^0 \pm x^3$,  $\vec x_\perp = (x^1, x^2)$, where $x^+$ is the light-front time, the direction of the dynamical evolution \cite{Brodsky:1997de}.
The ellipsis represents higher Fock sector contributions, which are also overlaps of LFWFs. At large $Q^2$, higher Fock sector contributions are suppressed, and (\ref{eqn:DYW}) reduces to the classical pQCD prediction \cite{Lepage:1980fj}, 
\begin{equation}\label{eqn:asymp}
F_\pi(Q^2) \xrightarrow{Q^2\to\infty} \frac{8\pi f_\pi^2\alpha_s(Q^2) }{Q^2}.
\end{equation}
Here, $f_\pi$ is the pion decay constant and $\alpha_s$ is the scale-dependent strong coupling constant. 

The Drell-Yan-West formula was derived in the Drell-Yan frame $q^+ = 0$, also known as the transverse frame. 
In principle, the form factor can be evaluated in frames other than the Drell-Yan frame. However, Isgur and Llewellyn-Smith showed that form factors evaluated with a longitudinal momentum transfer ($q^+\ne 0, \vec q_\perp = 0$) exhibit large discrepancies with form factors given by (\ref{eqn:DYW}) for available phenomenological wave functions \cite{Isgur:1988iw}. This apparent loss of covariance raised doubts about the Drell-Yan-West formula (\ref{eqn:DYW}) in the non-perturbative regime. 

\begin{figure}
\centering
\includegraphics[width=0.5\textwidth]{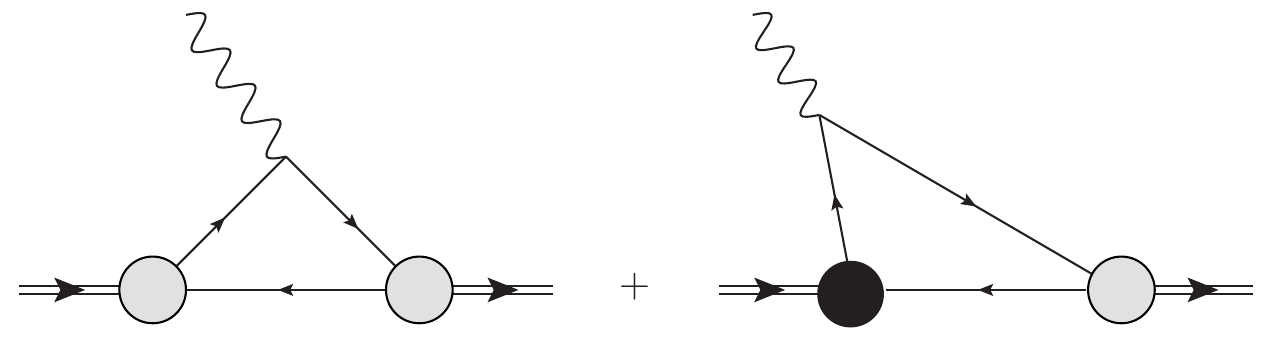}
\caption{(\textit{Left}) Overlap contributions for the form factor. (\textit{Right}) Non-overlap contributions for the form factor.
The shaded circles represent the light-front wave functions. The solid circles represent a non-wave function vertex. 
}
\label{fig:zero-mode}
\end{figure}

One of the possible ways to resolve this discrepancy is to start from the covariant Bethe-Salpeter equation \cite{Frankfurt:1977vc, Frankfurt:1981mk, Sawicki:1992qj, Demchuk:1995zx, deMelo:1997hh, deMelo:1997cb, deMelo:1998an, Choi:1998nf, deMelo:1999gn, Bakker:2000pk, Tiburzi:2001je, Melikhov:2001pm, Simula:2002vm, deMelo:2002yq, Tiburzi:2002mn, Melikhov:2002mp, deMelo:2003uk, Tiburzi:2004ye, deMelo:2005cy, Suzuki:2012ni, deMelo:2012hj}. After projecting the Bethe-Salpeter amplitude (BSA) on the light front in a general frame, apart from the overlap of wave functions described by the Drell-Yan-West formula, there emerges a new contribution, known as the $Z$-term or the pair (creation/annihilation) term \cite{Sawicki:1992qj, deMelo:1999gn, deMelo:2002yq}, as shown in Fig.~\ref{fig:zero-mode}. This term cannot be represented as the overlap of LFWFs, which deepens the doubts on the applicability of the light-front formalism to form factors in a general frame.

In principle, the $Z$-term can be obtained from the BSA.  
However, since the state-of-the-art hadronic BSAs are solved in Euclidean spacetime \cite{Eichmann:2016yit, Maris:2003vk, Carbonell:2017isq}, the access to form factors in a non-Drell-Yan frame is limited. 
Even the use of the Drell-Yan frame is questionable, because in this frame, the $Z$-term becomes vacuum pair creation and annihilation contributions and may not vanish as required by the Drell-Yan-West formula. It is well-known that modes with vanishing longitudinal momentum, i.e. zero modes, may populate the light-front vacuum, which may contribute to vacuum creation/annihilation processes in the Drell-Yan frame  \cite{deMelo:1997hh, deMelo:1997cb, deMelo:1998an, Choi:1998nf, deMelo:1999gn, Bakker:2000pk, Bakker:2002mt, Tiburzi:2001je, Melikhov:2001pm, Simula:2002vm, deMelo:2002yq, Tiburzi:2002mn, Melikhov:2002mp, deMelo:2003uk, Tiburzi:2004ye, Choi:2004ww, deMelo:2005cy, Choi:2005fj, He:2005hw, Suzuki:2012ni, deMelo:2012hj, Arifi:2022qnd}. However, it is useful to distinguish the $Z$-term and the dynamical zero-mode contributions, where the latter requires a dynamical treatment of the vacuum (see discussions in Sect.~\ref{sect:LF_Hamiltonian_formalism}). 

\begin{figure}
\centering
\includegraphics[width=0.25\textwidth]{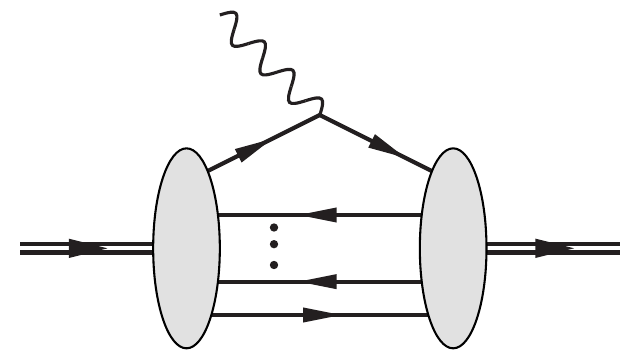} \quad
\includegraphics[width=0.25\textwidth]{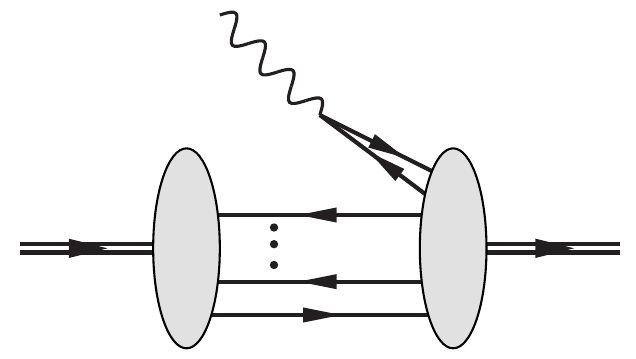}
\caption{(\textit{Left}) Diagonal contributions to the form factor: overlap of LFWFs with the same number of partons. (\textit{Right}) 
Non-diagonal contributions to the form factor: overlap of LFWFs with different numbers of partons. }
\label{fig:high_Fock}
\end{figure}

An alternative approach to resolve the issue of frame dependence is to incorporate higher Fock sector contributions directly on the light front \cite{Brodsky:1998hn}. In principle, wave functions in the full Fock space contain the complete information of the system. For example, the $Z$-term arises from the overlap of wave functions with different numbers of partons following the coupling of the current to the pair creation/annihilation vertex as shown in Fig.~\ref{fig:high_Fock}.  
The discrepancy between results obtained in different frames is caused by the Fock sector truncation used in practical calculations, which violates some of the Poincaré symmetries. 
Indeed, some of us argued that the frame dependence can be used as a figure of merit to gauge the violation of the Poincaré symmetry \cite{Li:2017uug, Li:2019kpr}. 

Using light cone perturbation theory (LCPT), Brodsky and Hwang showed that the zero modes do not contribute to the electromagnetic (e.m.) current in a scalar theory \cite{Brodsky:1998hn}. 
In non-perturbative calculations, limited by the computational power, one of the common strategies in the literature is to retain the valence Fock sector for the hadronic wave function, and incorporate the effects of the higher Fock sectors in the current operator. For example, 
in light front holography, the dressed current can be obtained from the mapping to the holographic current in AdS/QCD \cite{Brodsky:2007hb, Brodsky:2008pf, Li:2023izn}. 
The dressed current can be systematically constructed in the valence Fock space by considering constraints from Poincaré covariance, parity and charge conjugation symmetry, Hermiticity and current conservation  \cite{Lev:1994au, Lev:1998qz, Lev:1999me, Lev:2000vm, Lev:1999au, Pace:2001tzb, Pace:2006yj, deMelo:2006bs, Marinho:2007zzb, Pace:2008obp, Huang:2008jd, Pace:2011zz, Polyzou:2023ldi}.  

The increase of the computational power as well as the advance of theoretical techniques now allows us to solve 3+1D quantum field theories beyond the valence Fock sectors. Among various achievements, solution with three-body truncation was achieved in QED in discretized light cone quantization (DLCQ) \cite{Chabysheva:2009vm}. Solution with four-body truncation was achieved in scalar Yukawa theory using a Lagrange mesh \cite{Li:2014kfa, Li:2015iaw}. Wave functions with five-body truncation, incorporating both dynamical gluons and sea quarks, are recently obtained in QCD for the baryon sector within the framework of basis light front quantization (BLFQ)  \cite{Xu:2024sjt}. A review of some of the recent progress can found in Refs.~\cite{Hiller:2016itl, Gross:2022hyw}. These higher Fock sector wave functions allow us to address the frame dependence as well as the zero modes within the e.m. current directly.  

In this work, we investigate the Lorentz covariance of the charge form factor of a strongly coupled scalar by taking advantage of the recent non-perturbative solution of the scalar Yukawa theory with a three-body truncation. Our previous work for the same theory with a four-body truncation showed that the three-body truncation is numerically converged for a number of observables \cite{Li:2015iaw, Cao:2024fto, Duan:2024dhy}. Therefore, the three-body truncation suffices for our purpose. Our previous results were solved with the quenched approximation, i.e.~excluding the antiparticle degree of freedom.
However, without the antiparticle degree of freedom, the pair creation/annihilation contribution shown in Fig.~\ref{fig:high_Fock}, is absent. We therefore first solve the non-perturbative scalar Yukawa theory in the full (unquenched) three-body truncation, i.e. we add back the anti-particle degree of freedom in the Fock space. We note that this is the first time such a solution in full three-body truncation is obtained.  With this solution, we evaluate the form factor in  a general frame with $q^+ \ne 0$. We then investigate the zero-mode contribution to the current in the Drell-Yan limit $q^+ \to 0$. 
We also take advantage of the recent advance in the covariant analysis of light-front hadronic matrix elements (HMEs), which allows us to systematically investigate the current components and to extract the form factor without the contamination of spurious contributions due to the violation of the Poincaré symmetry \cite{Cao:2024rul}.  
 
The remainer of the work is organized as follows. We start with a brief introduction to the light-front Hamiltonian formalism in Sect.~\ref{sect:LF_Hamiltonian_formalism}. We then delve into the solution of the scalar Yukawa model with the full three-body truncation in Sect.~\ref{sect:threey-body_truncation}. Next, we analyze the HMEs of the e.m. current operator in covariant light front dynamics in Sect.~\ref{sect:EMFF}. From the covariant analysis, we compute the e.m. form factors, and analyze the zero-mode contributions. Finally, we conclude in Sect.~\ref{sect:summary}. 

\section{Light-front Hamiltonian formalism}\label{sect:LF_Hamiltonian_formalism}

In the Hamiltonian formalism, the dynamical evolution of a state vector is governed by Schrödinger equation \cite{Dirac:1949cp},
\begin{equation}\label{eqn:Schrödinger_equation}
i\frac{\partial}{\partial x^\mu} |\Psi(x) \rangle = P_\mu |\Psi(x) \rangle,
\end{equation}
where, $P_\mu$ is the 4-momentum operator. On the other hand, for a hadron with a momentum $p$, spin $j$ and magnetic projection $m_j$, Einstein's relativity requires that the hadronic state vector $| \psi_h(p, j, m_j) \rangle =e^{ip\cdot x} |\Psi(x) \rangle$ satisfies the relativistic dispersion relation \cite{Brodsky:1997de}, 
\begin{equation}\label{eqn:Einstein_equation}
 P_\mu P^\mu | \psi_h(p, j, m_j) \rangle = M^2_h | \psi_h(p, j, m_j) \rangle,
\end{equation}
where, $M_h$ is the invariant mass of the hadron $h$. The invariant mass squared operator $H \equiv  P_\mu P^\mu$ is also known as the light cone Hamiltonian. 
In light-front dynamics, the system evolves in the direction of the light-front time $x^+$, and the invariant mass squared operator can be written as $H = P^+ P^- - \vec P_\perp^2$. 
Since $P^+$ and $P_\perp$ are purely kinematical, for the hadronic state, they can be taken as the 3-momentum of the hadron, i.e. 
$H = p^+P^- - p^2_\perp$. Consequently, $H$ is a linear operator of the Hamiltonian operator $P^-$, and Einstein equation (\ref{eqn:Einstein_equation}) is strictly equivalent to the Schrödinger equation (\ref{eqn:Schrödinger_equation}) for all momentum $(p^+, \vec p_\perp)$. 
In this way, wave functions on the light front are known to be invariant under the following light-front boost transformations,\cite{Kogut:1969xa, Brodsky:1997de}
\begin{equation}\label{eqn:boosts}
\begin{split}
\text{longitudinal boosts } K^3 \equiv \frac{1}{2}M^{+-}: \; p^+ \; \to & \; p'^+ = e^{\phi}p^+, \qquad \vec p_\perp \; \to\; \vec p'_\perp = \vec p_\perp, \\
\text{transverse boosts } \vec B_\perp \equiv M^{+i}: \; p^+ \;\to&\; p'^+ = p^+, \qquad \vec p_\perp \; \to \; \vec p'_\perp = \vec p_\perp + p^+\vec \beta_\perp.
\end{split}
\end{equation} 
where, $M^{\mu\nu}$ are the generators of the Poincaré algebra.

In the light-front Hamiltonian formalism, the main task is to solve the eigenvalue equation (\ref{eqn:Einstein_equation}),
\begin{equation}\label{eqn:light-cone_eigenvalue_equation}
 H | \psi_h(p, j, m_j) \rangle = M^2_h | \psi_h(p, j, m_j) \rangle.
\end{equation}
In DLCQ \cite{Pauli:1985pv, Pauli:1985ps} and BLFQ \cite{Vary:2009gt}, a set of finite many-body basis is chosen to render Eq.~(\ref{eqn:light-cone_eigenvalue_equation}) a matrix eigenvalue equation. The standard numerical eigensolvers can be used to obtain the mass eigenvalues $M_h$ and the corresponding wave functions. The difference between DLQC and BLFQ is the choice of the basis: the latter chooses a basis that preserving all kinematical symmetries of the light cone Hamiltonian $H$. In this work, we adopt an alternative method. We first impose a Fock sector truncation on the Hilbert space and obtain a set of coupled integral equations. We then approximate the integrals using Gauss quadratures \cite{Brodsky:2005yu}. This method, known as the Lagrange mesh method,  is shown to be equivalent to adopting a basis of (weighted) orthogonal polynomials \cite{Baye:2015xoi, Baye:2002tix}. 

The direction of the dynamical evolution classifies the operators into two categories: kinematical operators and dynamical operators \cite{Dirac:1949cp, Leutwyler:1977vy}. Kinematical operators preserve the initial surface while dynamical operators do not. 
Hence, dynamical operators must contain the interaction. Light-front dynamics turns out to be equipped with the maximal number (7) of kinematical operators out of the 10 Poincaré generators. The remaining 3 generators originate from the same local interaction in quantum field theory \cite{Carbonell:1998rj}, 
\begin{align}
P^\mu_\text{int} =\,& \omega^\mu \int \dd^4 x\, \delta(\omega\cdot x) \mathcal H_\text{int}(x), \\
J^{\mu\nu}_\text{int} =\,& \int \dd^4 x\, \delta(\omega\cdot x)  (x^\mu \omega^\nu - \omega^\mu x^\nu) \mathcal H_\text{int}(x)\,.
\end{align}
Here, $\omega^\mu = (\omega^+, \omega^-, \vec\omega_\perp) = (0, 2, \vec 0_\perp)$ is the null vector that is perpendicular to the orientation of the light-front quantization surface $\Sigma = \{ x \in \mathbb R^{1,3} | \omega_\mu x^\mu = 0\}$. 
Besides the light-front energy $P^-$, generators of the transverse rotations are also dynamical. The dynamical nature of these operators means that they are likely broken when a finite Fock sector truncation is adopted. This is the origin of the violation of Poincaré symmetries in practical calculations in light-front dynamics.
The light-front boosts are kinematical. Furthermore, transverse boosts on the light front are Galilean, as shown in Eq.~(\ref{eqn:boosts}).  These features have motived a rigorous definition of an intrinsic hadronic charge density as \cite{Miller:2007uy, Miller:2010nz}, 
\begin{align}
\rho_\textsc{lf}(r_\perp) =\,& \int \frac{\dd q^+\dd^2 q_\perp}{(2\pi)^32P^+} e^{-i\vec q_\perp \cdot \vec r_\perp} \langle P+\frac{1}{2}q | \Sigma(0) |P-\frac{1}{2}q \rangle, \\
=\,&  \int \frac{\dd^2 q_\perp}{(2\pi)^2} e^{-i\vec q_\perp \cdot \vec r_\perp} F(q^2_\perp)
\end{align}
where, the transverse density operator is defined as the light-front projection of the light-front charge density,
\begin{equation}
\Sigma(\vec x_\perp) = \frac{1}{2} \int \dd x^- J^+(x^-, \vec x_\perp; x^+=0).
\end{equation}
 Unlike the standard Sachs density defined in equal-time quantization, the light-front density is exact and independent of the choice of $P^+$ and $\vec P_\perp$. Note that light-front density can be directly obtained from the 2D Fourier transform of the form factor within the Drell-Yan frame $q^+=0$, which highlights the special role played by the Drell-Yan frame in light-front dynamics \cite{Miller:2007uy, Miller:2009sg, Miller:2009qu, Miller:2010nz,  Miller:2018ybm,  Freese:2021czn, Freese:2021qtb, Freese:2021mzg, Freese:2022fat, Li:2022hyf, Freese:2023jcp, Freese:2023abr}. 

The hadronic state vector can be represented in Fock space, which consists of an infinite number of non-interacting particles. In Fock space, particles are on their mass shell. Hence light-front energy conservation is not held at the interaction vertices. 
The on-shellness of Fock space particles combined with the positivity of the light-front energy $p^-_i$ leads to one of the most striking features of light-front dynamics: 
\begin{equation}\label{eqn:spectra_condition}
p^+_i = \frac{p^2_{i\perp} + m_i^2}{p^-_i} \ge 0.
\end{equation}
Eq.~(\ref{eqn:spectra_condition}) is known as the spectral condition. Because the physical vacuum is defined as a state with a vanishing longitudinal momentum $p^+ = 0$, the spectra condition seems to imply that the physical vacuum contains no particles, i.e. it is simply the Fock vacuum \cite{Brodsky:2009zd}. 

The above argument has a loophole. Modes with vanishing longitudinal momentum $p_i^+ = 0$ can populate the vacuum. Indeed, zero modes are responsible for physics arising from a non-trivial vacuum, e.g. spontaneous symmetry breaking  \cite{Hornbostel:1991qj, Heinzl:1991vd, Chakrabarti:2003tc, Martinovic:2002bv}. 
Careful analysis based on the light-front projection of the covariant amplitudes shows that zero modes contribute to the vacuum amplitudes \cite{Chang:1968bh, Yan:1973qg, Chabysheva:2022duu}, pair production processes \cite{Ilderton:2014mla, Tomaras:2001vs}, light-front effective potentials \cite{Heinzl:2002uy}, the $Z$-term of the form factor \cite{deMelo:1998an}, and so on. On the light-front Hamiltonian side, since the light-front energy of the zero modes diverges $({p^2_{i\perp} + m_i^2})/{p^+_i} \to \infty$, it has not been clear for a long time how to properly incorporate them in practical calculations or to determine whether they truly contribute to various observables \cite{Burkardt:1995ct, Burkardt:2000jn}. 
In discretized light-front field theory, all modes are discretized and the zero modes can be separated from the normal modes. In scalar theory, the zero modes are relatively simple -- they are shown to be a constrained degree of freedom \cite{McCartor:1991ch, Robertson:1992nj, Pinsky:1993yi, Heinzl:1995xj}. 

In a continuum approach like our work \cite{Perry:1990mz}, the zero modes cannot be separated \cite{Burkardt:1992sz, Fitzpatrick:2018ttk}. Recently, Collins proposed that the vacuum diagrams can be properly evaluated on the light front by taking the limit of vanishing external momentum \cite{Collins:2018aqt, Martinovic:2018apr}. This method allows us to work directly in light-front field theory, and the zero modes are automatically incorporated if the endpoints are properly treated. 
We apply this method to investigate the zero modes within the e.m. current. We compute the form factor in a general frame with $q^+ \ne 0$ and  take the zero photon momentum limit $q^+ \to 0$ to explore the possible zero-mode contribution within the form factor.

\section{Non-perturbative solution of scalar Yukawa theory within three-body truncation}\label{sect:threey-body_truncation}

\subsection{Scalar Yukawa theory}\label{sect:scalar_Yukawa_theory}

The Lagrangian of the scalar Yukawa model is,
\begin{equation}
\mathscr L = \partial_\mu N^\dagger\partial^\mu N - m^2  |N |^2 + \frac{1}{2} \partial^\mu \pi\partial_\mu \pi - \frac{1}{2}\mu^2 \pi^2 + g   |N |^2 \pi.
\end{equation}
where, $ N$ presents the (mock) nucleon field and $\pi$ the (mock) pion field. We thus tentatively assign $m = 0.94\,\mathrm{GeV}$ the nucleon mass and $\mu = 0.14\,\mathrm{GeV}$ the pion mass. The coupling $g$ is dimensional, and it is useful to introduce a dimensionless coupling: $\alpha = g^2/(16\pi m^2)$, which is the coupling of the tree-level Yukawa potential, 
\begin{equation}
V(r) = -\frac{\alpha e^{-\mu r}}{r}.
\end{equation} 

While the theory is super-renormalizable, it still possesses divergences. We adopt the Pauli-Villars (PV) regularization scheme \cite{Brodsky:1998hs}. We introduce two types of PV particles, one PV particle accompanying each species. The regularized Lagrangian is, 
\begin{multline}
\mathscr L = \sum_{i=0}^1 (-1)^i \big[ \partial_\mu  N^\dagger_i\partial^\mu N_i - m^2_i N^\dagger_i N_i \big] + \frac{1}{2} \sum_{j=0}^1(-1)^j  \big[ \partial^\mu \pi_j\partial_\mu \pi_j - \frac{1}{2}\mu^2_j \pi^2_j \big] + \sum_{i, i', j=0}^1  g_0  N^\dagger_i  N_{i'} \pi_j \\
+ \sum_{i=0}^1 (-1)^i \delta m_i^2  N^\dagger_i  N_i + \frac{1}{2}\sum_{j=0}^1 (-1)^j \delta \mu_j^2 \pi_j^2,
\end{multline}
where, $g_0$ is the bare coupling,  $\delta m^2_i = m^2_i - m_{0i}^2$ $\delta \mu^2_j \equiv \mu^2_j - \mu_{0j}^2$ are the mass counterterms. Subscript ``0" indicates the physical quantities, e.g.~the physical masses $m_0 \equiv m$ and $\mu_0 \equiv \mu$, and subscript ``1" indicates the PV quantities, e.g. the PV masses $m_1 \equiv m_\textsc{pv}$ and $\mu_1 \equiv \mu_\textsc{pv}$. We choose $0 < \mu < m \ll \mu_\textsc{pv} < m_\textsc{pv}$.

Previously, the scalar theory is solved with a Fock sector expansion up to three particles ($|N\rangle + |\pi N\rangle + |\pi\pi N\rangle$) \cite{Karmanov:2008br, Karmanov:2016yzu} and four particles ($| N\rangle + |\pi N\rangle + |\pi\pi N\rangle + |\pi\pi\pi N\rangle$) \cite{Li:2014kfa, Li:2015iaw}. By comparing solutions from the three-body and the four-body truncations, the Fock sector convergence is numerically achieved even in the non-perturbative regime.  
Note that these solutions do not include anti-particle degrees of freedom within the Fock space. 
With the quenched solutions, the non-diagonal contributions, i.e. the pair creation/annihilation terms, are not present in frames with $q^+ \ne 0$. 
As we will show later in Sect.~\ref{sect:frame_dependence}, there is a large frame dependence for the form factor of the quenched solutions, as exemplified by the discrepancy between results from two representative frames,  the transverse frame (Drell-Yan frame) $q^+=0$ and the longitudinal frame $\vec\Delta_\perp = \vec q_\perp - q^+ (\vec p_\perp+\vec p'_\perp)/(p^++p'^+) = 0$. This large frame dependence is the consequence of the violation of the Poincaré symmetry. 
Karmanov~et.~al. showed that, in the context of the Yukawa model, the Poincaré symmetry can be partially restored by incorporating the anti-particle degree of freedom \cite{Karmanov:2012aj}.  
In this work, we first extend the solution to the full three-body Fock sector, 
\begin{equation}
| N\rangle_\text{ph} = | N\rangle + | \pi N\rangle + | \pi\pi N\rangle + |N N\bar N\rangle.
\end{equation}
Namely, the anti-nucleon degree of freedom now is incorporated. We also refer to this solution as the unquenched solution. 

\subsection{Eigenvalue equation}\label{sect:eigenvalue_equation}

\begin{figure}
\centering
\includegraphics[width=0.75\textwidth]{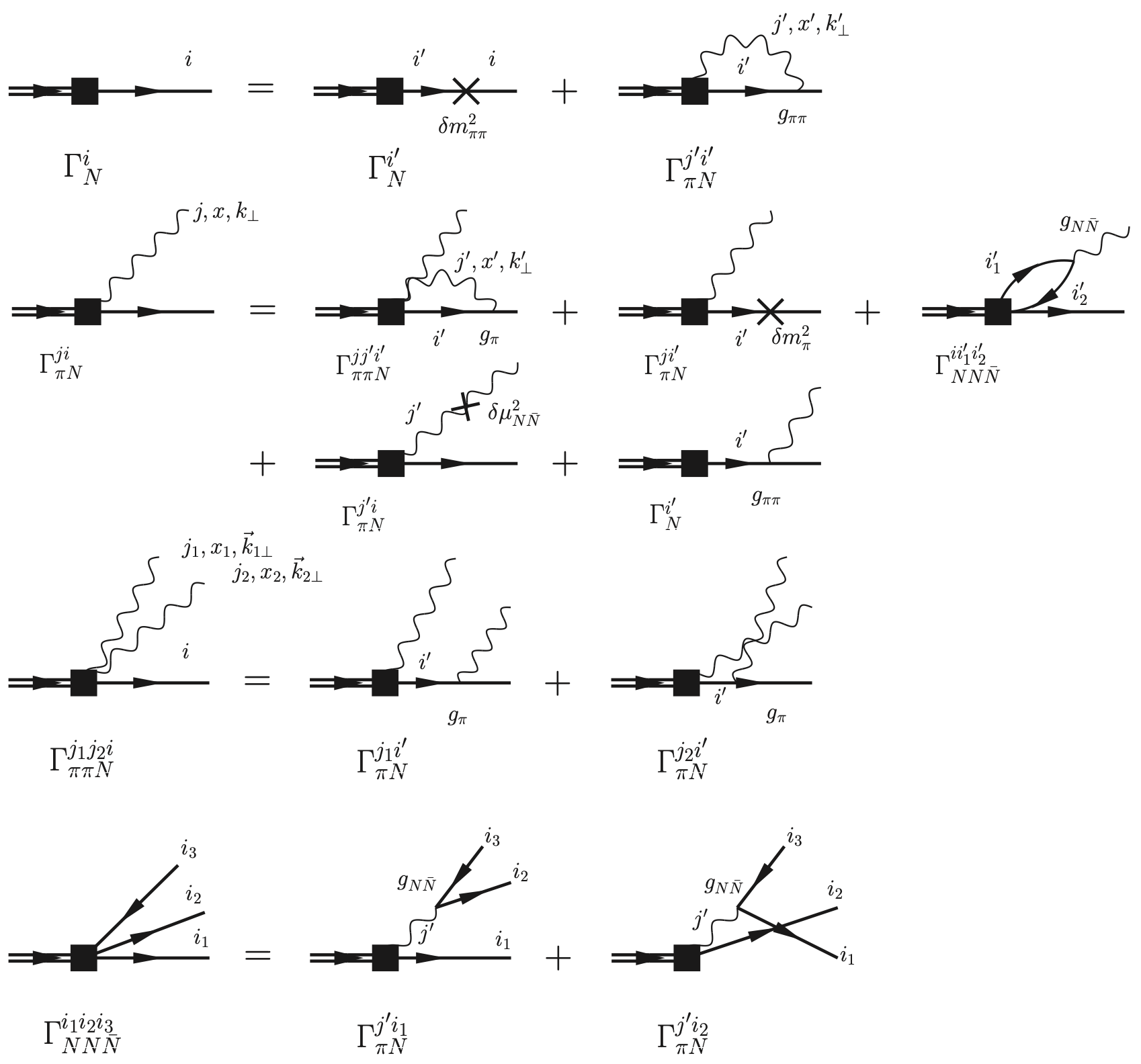}
\caption{Diagrammatic representation of the system of equations (\ref{eqn:Gamma_chi})--(\ref{eqn:Gamma_chi_chi_barchi}) for the eigenvalue equation within three-body Fock sector truncation. The double lines represent the physical state, i.e. the mock physical nucleon. The solid lines represent the charged scalar, i.e. the mock nucleon or anti-nucleon. The wavy lines represent the real scalar, i.e. the mock pion. The shaded boxes represent the vertex functions $\Gamma$. The bare parameters are assigned according to sector dependent scheme. }
\label{fig:SOE}
\end{figure}

The diagrammatic representation of the system of equations\footnote{These system of equations are the light-front version of the Kadyshevsky equation -- relativistic generalization of the Lippmann-Schwinger equation, which is directly equivalent to the eigenvalue equation (\ref{eqn:light-cone_eigenvalue_equation}). See Ref.~\cite{Carbonell:1998rj} for a review. } is shown in Fig.~\ref{fig:SOE}. We have assigned counterterms (bare coupling and mass counterterms) the symbols according to the available Fock sector as specified by Fock sector dependence renormalization. According to this scheme, $g_\pi$, $\delta m_\pi^2$, $g_{ N\bar N}$ and $\delta \mu^2_{ N\bar N}$ can be obtained from lower Fock sector truncations, i.e. $|N\rangle+|\pi N\rangle$, and $|\pi\rangle+|N\bar N\rangle$, which are summarized in the Appendix~\ref{sect:FSDR_two-body}. Then, $g_{\pi\pi}$ and $\delta m^2_{\pi\pi}$ are the only counterterms that need to be determined. We will discuss the normalization of these parameters in Sect.~\ref{sect:FSDR}.  

Employing the light-font Feynman rules\footnote{These are sometimes known as the Weinberg rules for the perturbative case \cite{Weinberg:1966jm} and Kadyshevsky rules for the bound states \cite{Kadyshevsky:1967rs, Atakishiev:1976sa, Karmanov:1976iv}.} \cite{Karmanov:2008br}, we can write down the system of equations for each vertex. 
The one-body vertex $\Gamma_ N$ satisfies, 
\begin{equation}
\label{eqn:Gamma_chi}
\Gamma_ N^i = \delta m_{\pi\pi}^2 \sum_{i'}(-1)^{i'} \frac{\Gamma_ N^{i'}}{m^2_{i'} - M^2}
+ \sum_{i', j'}(-1)^{i'+j'} \int\limits_0^1\frac{\dd x'}{2x'(1-x')}\int\frac{\dd^2k'_\perp}{(2\pi)^3} \frac{g_{\pi\pi}\Gamma_{\pi N}^{j'i'}(x', k'_\perp)}{s_2^{j'i'} - M^2}.
\end{equation}
Here $M \to m$ is the mass eigenvalue and  
\begin{equation}
s_2^{j'i'} = \frac{k'^2_\perp+\mu_{j'}^2}{x'} + \frac{k'^2_\perp+m_{i'}^2}{1-x'}.
\end{equation}
The two-body vertex $\Gamma_{\pi N}$ satisfies,  
\begin{equation}\label{eqn:Gamma_chi_varphi}
\begin{split}
\Gamma^{ji}_{\pi N}(x, k_\perp) =\,& 
\sum_{i',j'} (-1)^{i'+j'} \int\limits_0^{1-x}\frac{\dd x'}{2x'(1-x-x')}\int\frac{\dd^2k'_\perp}{(2\pi)^3} \frac{g_\pi \Gamma_{\pi\pi N}^{jj'i'}(x, \vec k_\perp, x', \vec k'_\perp)}{s_{3a}^{jj'i'} - M^2} \\
& + \sum_{i'}(-1)^{i'} \frac{\delta m^2_\pi}{1-x} \frac{\Gamma_{\pi N}^{ji'}(x, k_\perp)}{s_{2a}^{ji'}-M^2} \\
& + \sum_{i'_1, i'_2}(-1)^{i'_1+i'_2}\int\limits_0^x\frac{\dd x'}{2x'(x-x')}\int\frac{\dd^2k'_\perp}{(2\pi)^3} \frac{g_{ N\bar N}\Gamma_{ N N\bar N}^{ii'_1i'_2}(1-x, -\vec k_\perp, x', \vec k'_\perp)}{s_{3b}^{ii'_1i'_2}-M^2} \\
& + \sum_{j'}(-1)^{j'} \frac{\delta \mu^2_{ N\bar N}}{x} \frac{\Gamma_{\pi N}^{j'i}(x,  k_\perp)}{s_{2b}^{j'i}-M^2} \\
& + \sum_{i'}(-1)^{i'}g_{\pi\pi}\frac{\Gamma_ N^{i'}}{m^2_{i'}-M^2}. 
\end{split}
\end{equation}
Here, 
\begin{align}
s_{3a}^{jj'i'} =\, & \frac{k_\perp^2+\mu_j^2}{x} + \frac{k'^2_\perp+\mu^2_{j'}}{x'} + \frac{(\vec k_\perp+\vec k'_\perp)^2+m^2_{i'}}{1-x-x'}, \\
s_{2a}^{ji'} =\,& \frac{k_\perp^2+\mu^2_j}{x} + \frac{k_\perp^2+m^2_{i'}}{1-x}, \\
s_{3b}^{ii'_1i'_2}=\,& \frac{k_\perp^2+m_i^2}{1-x} + \frac{k'^2_\perp+m^2_{i'_1}}{x'} + \frac{(\vec k_\perp-\vec k'_\perp)^2+m^2_{i'_2}}{x-x'}, \\
s_{2b}^{j'i} =\,& \frac{k_\perp^2+\mu^2_{j'}}{x} + \frac{k_\perp^2+m^2_i}{1-x} 
\end{align}
The three-body vertex $\Gamma_{\pi\pi N}$ satisfies, 
\begin{equation}\label{eqn:Gamma_chi_varphi_varphi}
\Gamma_{\pi\pi N}^{j_1j_2i}(x_1, \vec k_{1\perp}, x_2, \vec k_{2\perp}) = 
\sum_{i'}(-1)^{i'} \frac{g_\pi}{1-x_1}\frac{\Gamma_{\pi N}^{j_1i'}(x_1, k_{1\perp})}{s_{2a}^{j_1i'}-M^2}
+ 
\sum_{i'}(-1)^{i'} \frac{g_\pi}{1-x_2}\frac{\Gamma_{\pi N}^{j_2i'}(x_2, k_{2\perp})}{s_{2b}^{j_2i'}-M^2}
\end{equation}
where 
\begin{align}
s_{2a}^{j_1i'} = \frac{k_{1\perp}^2+\mu^2_{j_1}}{x_1} + \frac{k_{1\perp}^2+m_{i'}^2}{1-x_1}, \\
s_{2b}^{j_2i'} = \frac{k_{2\perp}^2+\mu^2_{j_2}}{x_2} + \frac{k_{2\perp}^2+m_{i'}^2}{1-x_2}.
\end{align}
The three-body vertex $\Gamma_{ N N\bar N}$ satisifies,  
\begin{multline}\label{eqn:Gamma_chi_chi_barchi}
\Gamma_{ N N\bar N}^{i_1i_2i_3}(x_1, \vec k_{1\perp}, x_2, \vec k_{2\perp}) = \sum_{j'}(-1)^{j'} \frac{g_{ N\bar N}}{1-x_1} \frac{\Gamma_{\pi N}^{j'i_1}(1-x_1, k_{1\perp})}{s_{2a}^{j'i_1}-M^2} \\
+ \sum_{j'}(-1)^{j'}\frac{g_{ N\bar N}}{1-x_2}\frac{\Gamma_{\pi N}^{j'i_2}(1-x_2, k_{2\perp})}{s_{2b}^{j'i_2}-M^2}.
\end{multline}
Here 
\begin{align}
s_{2a}^{j'i_1} =\,& \frac{k_{1\perp}^2+\mu^2_{j'}}{1-x_1} + \frac{k_{1\perp}^2+m^2_{i_1}}{x_1}, \\
s_{2b}^{j'i_2} =\,& \frac{k_{2\perp}^2+\mu^2_{j'}}{1-x_2} + \frac{k_{2\perp}^2+m^2_{i_2}}{x_2}.
\end{align}
The Fock sector truncation provides a closure to the system of equations.

\subsection{Cluster reduction}

\begin{figure}
\centering
\includegraphics[width=0.75\textwidth]{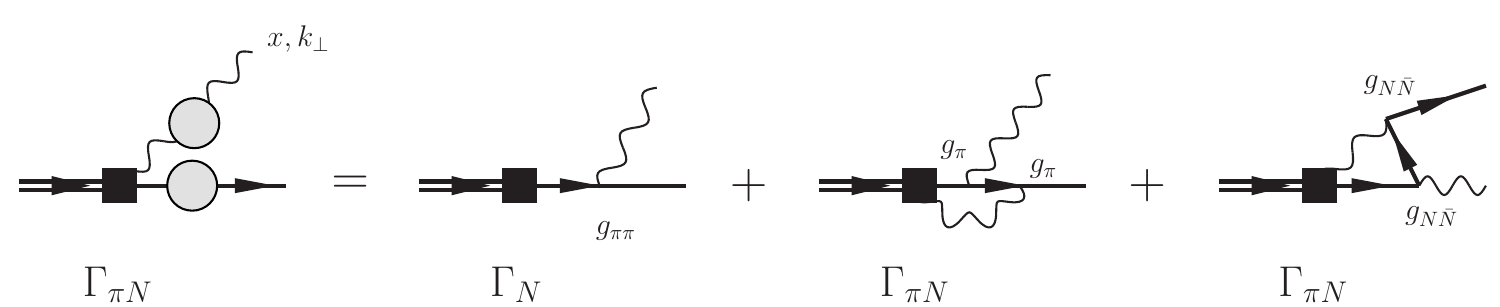}
\caption{The diagrammatic representation of Eq.~(\ref{eqn:Gamma_chi_varphi_c}). 
We have suppressed the Pauli-Villars indices for simplicity. The shaded circles represent self-energies. See Fig.~\ref{fig:SOE} for the interpretation of the remaining symbols.}
\label{fig:Gamma_chi_varphi_cc}
\end{figure}

To simplify the system of equations,  we can express
the three-body vertices $\Gamma_{\pi\pi N}$ and $\Gamma_{ N N\bar N}$ 
as a function of $\Gamma_{\pi N}$ and obtain a new equation that only involves the two-body vertex $\Gamma_{\pi N}$ as shown in Fig.~\ref{fig:Gamma_chi_varphi_cc}. The relevant expression is, 
\begin{equation}\label{eqn:Gamma_chi_varphi_c}
\begin{split}
\bigg[ 
& \Gamma_{\pi N}^{ji}(x, k_\perp) -
\sum_{i'}(-1)^{i'} \frac{\Sigma_\pi(t_a^j)-\Sigma_\pi(m^2)}{t_a^j- m^2_{i'}} 
\Gamma_{\pi N}^{ji'}(x, k_\perp) -
\sum_{j'}(-1)^{j'} \frac{\Pi_{ N\bar N}(t_b^i)-\Pi_{ N\bar N}(\mu^2)}{t_b^i-\mu^2_{j'}} 
\Gamma_{\pi N}^{j'i}(x, k_\perp) \bigg]
\\
=\,& \sum_{i'}(-1)^{i'}g_{\pi\pi}\frac{\Gamma_ N^{i'}}{m^2_{i'}-M^2} \\
 & + g_\pi^2 \sum_{i', j'} (-1)^{i'+j'} \int\limits_0^{1-x} \frac{\dd x'}{2x'(1-x')(1-x-x')}
\int\frac{\dd^2k'_\perp}{(2\pi)^3} \frac{\Gamma_{\pi N}^{j'i'}(x', k'_\perp)}{s_{2c}^{j'i'}-M^2} 
\sum_{i''} \frac{(-1)^{i''}}{s_{3c}^{jj'i''}-M^2}  \\
& + g_{ N\bar N}^2 \sum_{i', j'} (-1)^{i'+j'} \int\limits_{1-x}^1 \frac{\dd x'}{2x'(1-x')(x+x'-1)}\int\frac{\dd^2k'_\perp}{(2\pi)^3} \frac{\Gamma_{\pi N}^{j'i'}(x', k'_\perp)}{s_{2d}^{j'i'}-M^2} 
\sum_{i''}\frac{(-1)^{i''}}{s_{3d}^{ii'i''}-M^2} \\
\end{split}
\end{equation}
where,  $t_a^j =  m^2_{i'} - (1-x)\big(s_{2a}^{ji'} - M^2\big)$, $t_b^i = \mu^2_{j''} - x\big( s_{2b}^{j''i} - M^2 \big)$, and,
\begin{align}
s_{2a}^{ji''} =\,& \frac{k_\perp^2+\mu^2_{j}}{x} + \frac{k_\perp^2+m_{i''}^2}{1-x}, \\
s_{2b}^{j''i} =\,& \frac{k^2_\perp+\mu^2_{j''}}{x} + \frac{k^2_\perp+m_{i}^2}{1-x}, \\
s_{2c}^{j''i''} =\,& \frac{k''^2_\perp+\mu^2_{j''}}{x''} + \frac{k''^2_\perp+m^2_{i''}}{1-x''}, \\
s_{3c}^{jj''i'} =\,& \frac{k_\perp^2+\mu_j^2}{x} + \frac{k''^2_\perp+\mu_{j''}^2}{x''} + \frac{(\vec k_\perp+\vec k''_\perp)^2+m^2_{i'}}{1-x-x''}, \\
s_{2d}^{j''i''} =\,& \frac{k''^2_\perp+\mu^2_{j''}}{x''} + \frac{k''^2_\perp+m^2_{i''}}{1-x''}, \\
s_{3d}^{ii'i''}=\,& \frac{k_\perp^2+m_i^2}{1-x} + \frac{k''^2_\perp+m^2_{i''}}{1-x''} + \frac{(\vec k''_\perp+\vec k_\perp)^2+m^2_{i'}}{x+x''-1}.
\end{align}
We have introduced the two-body self-energies:
\begin{align}
\Sigma_{\pi}(p^2) =\,& - g_\pi^2 \sum_{i', j'} (-1)^{i'+j'}\int_0^1 \frac{\dd x'}{2x'(1-x')}\int\frac{\dd^2k'_\perp}{(2\pi)^3}\frac{1}{s_2^{j'i'}-p^2}. \\
\Pi_{ N\bar N}(k^2) =\,& -g_{ N\bar N}^2\sum_{i_1', i'_2} (-1)^{i'_1+i'_2} \int_0^1 \frac{\dd x'}{2x'(1-x')}
\int\frac{\dd^2 k'_\perp}{(2\pi)^3}\frac{1}{s_2^{i'_1i'_2}-k^2}. 
\end{align}
Note that, $\delta m^2_\pi = \Sigma_\pi(m^2)$, and $\delta \mu^2 = \Sigma_{ N\bar N}(\mu^2)$. The two-body self-energies are computed analytically in the Appendix. 
Equation~(\ref{eqn:Gamma_chi_varphi_c}) is nothing but the light-front Hamiltonian version of the Dyson-Schwinger equations. 
In Hamiltonian dynamics, all diagrams are time ordered. For example, the second and third diagrams on the r.h.s. of the equation in Fig.~\ref{fig:Gamma_chi_varphi_cc} are the same covariant diagram. 
Note that the third diagram involves the anti-nucleon degree of freedom, which signifies the importance of the anti-particle degree of freedom in maintaining the Lorentz covariance. 

The scalar theory is super-renormalizable. The only divergence appears at the one-loop level, e.g., 
$\Sigma_\pi$ and $\Pi_{ N\bar N}$. The vertex functions as well as the $Z$ factors 
do not contain further divergences. As such, the regulators can be removed 
after the mass counter-term subtraction:
\begin{align}
\Sigma^R_\pi(p^2) \equiv\,& \Sigma_\pi(t) - \Sigma_\pi(m^2) \nonumber \\
=\,& \lim_{m_\textsc{pv}\to\infty} \lim_{\mu_\textsc{pv} \to \infty} g_\pi^2 \sum_{i', j'} (-1)^{i'+j'}\int_0^1 \frac{\dd x'}{2x'(1-x')}\int\frac{\dd^2k'_\perp}{(2\pi)^3} \Big( \frac{1}{s_2^{j'i'}-m^2} - \frac{1}{s_2^{j'i'}-p^2} \Big)  \nonumber \\
=\,& g_\pi^2 \int_0^1 \frac{\dd x'}{2x'(1-x')}\int\frac{\dd^2k'_\perp}{(2\pi)^3} \Big( \frac{1}{s_2-m^2} - \frac{1}{s_2-p^2} \Big);  \\
\Pi_{ N\bar N}^R(k^2) \equiv\,& \Pi_{ N\bar N}(k^2)  - \Pi_{ N\bar N}(\mu ^2)  \nonumber\\
=\,&\lim_{m_\textsc{pv}\to\infty} \lim_{\mu_\textsc{pv} \to \infty}  g_{ N\bar N}^2\sum_{i_1', i'_2} (-1)^{i'_1+i'_2} \int_0^1 \frac{\dd x'}{2x'(1-x')}
\int\frac{\dd^2 k'_\perp}{(2\pi)^3}\Big( \frac{1}{s_2^{i'_1i'_2}-\mu^2} -  \frac{1}{s_2^{i'_1i'_2}-k^2} \Big) \nonumber \\
=\,& g_{ N\bar N}^2 \int_0^1 \frac{\dd x'}{2x'(1-x')}
\int\frac{\dd^2 k'_\perp}{(2\pi)^3}\Big( \frac{1}{s_2-\mu^2} -  \frac{1}{s_2-k^2} \Big). 
\end{align}
Equation~(\ref{eqn:Gamma_chi_varphi_c}) may be written as,
\begin{equation}\label{eqn:Gamma_chi_varphi_cc}
\begin{split}
 \bigg[ 
 1 -
\frac{\Sigma_\pi^R(t_a)}{t_a - m^2}   - \,&
\frac{\Pi_{ N\bar N}^R(t_b)}{t_b-\mu^2}  \bigg]
\Gamma_{\pi N}(x, k_\perp)
\\
= g_{\pi\pi}\frac{\Gamma_ N}{m^2-M^2}
  +\,&  g_\pi^2 \int\limits_0^{1-x} \frac{\dd x'}{2x'(1-x')(1-x-x')}
\int\frac{\dd^2k'_\perp}{(2\pi)^3} \frac{\Gamma_{\pi N}(x', k'_\perp)}{s'_{2}-M^2} 
\frac{1}{s_{3c}-M^2}  \\
+&\, g_{ N\bar N}^2 \int\limits_{1-x}^1 \frac{\dd x'}{2x'(1-x')(x+x'-1)}\int\frac{\dd^2k'_\perp}{(2\pi)^3} \frac{\Gamma_{\pi N}(x', k'_\perp)}{s'_{2}-M^2} 
\frac{1}{s_{3d}-M^2}. \\
\end{split}
\end{equation}
The symbols are the same as Eq.~(\ref{eqn:Gamma_chi_varphi_c}), except only the physical parts remain. 
Equation~(\ref{eqn:Gamma_chi_varphi_cc}) is the main equation to be solved in this section. 

We may identify that, 
\begin{align}
Z_2(t) =\,&  \bigg( 1 - \frac{\Sigma^R(t)}{t - m^2} \bigg)^{-1}, \\
Z_3(t) =\,& \bigg( 1 - \frac{\Pi^R(t)}{t - \mu^2} \bigg)^{-1},
\end{align}
as the off-shell $Z$-factor. The on-shell $Z$-factors are obtained by taking the respective on-shell limit, 
\begin{align}
Z_2 \equiv\,& \lim_{t \to m^2} Z_2(t) = \bigg( 1 - \frac{\dd }{\dd t} \Sigma(t)\bigg)^{-1}_{t\to m^2}, \\
Z_3 \equiv\,& \lim_{t \to m^2} Z_3(t) = \bigg( 1 - \frac{\dd }{\dd t} \Pi(t)\bigg)^{-1}_{t\to \mu^2}.
\end{align}
While these definitions hold for the general Fock sector expansion, in Eq.~(\ref{eqn:Gamma_chi_varphi_cc}), the $Z$-factors are the two-body approximated versions due to Fock sector truncation, i.e. $Z_2 = Z_\pi, Z_3 = Z_{N\bar N}$, where $Z_\pi, Z_{N\bar N}$ are given in the Appendix. 
Note that in Eq.~(\ref{eqn:Gamma_chi_varphi_cc}), both legs are dressed (see l.h.s. of the equation in Fig.~\ref{fig:Gamma_chi_varphi_cc}).  However, the relevant off-shell $Z$-factor is not the product of $Z_2(t_a)$ and $Z_3(t_b)$. Instead, it is, 
\begin{equation}\label{eqn:hybridZ}
Z_{23}(s) =  \bigg[ 
 1 -
\frac{\Sigma^R(t_a)}{t_a - m^2}  -
\frac{\Pi^R(t_b)}{t_b-\mu^2}  \bigg]^{-1} \ne Z_2(t_a) Z_3(t_b),
\end{equation}
where, $t_a = m^2 - (1-x)(s-M^2)$, and $t_b = \mu^2 - x(s-M^2)$. 
The difference arises from the fact that the product $Z_2(t_a) Z_3(t_b)$, while consistent with the cluster decomposition,
 is not consistent with the three-body Fock sector truncation. On the mass shell $s = m^2$, $Z_{23}$ becomes, 
 \begin{align}
 Z_{23} =\,& \lim_{s \to m^2} \Big(1 - (1 - Z_2^{-1}) - (1 - Z_3^{-1}) \Big)^{-1}, \\
 =\,& \frac{Z_2Z_3}{Z_2+Z_3-Z_2Z_3}, \\
 =\,&  \frac{Z_\pi Z_{N\bar N}}{Z_\pi+Z_{N\bar N}-Z_\pi Z_{N\bar N}}.
 \end{align}
 In the last line, we use the two-body approximation (see Appendix) to replace the on-shell Z-factors, i.e. $Z_2 \to Z_\pi$ and $Z_{N\bar N}$.

\subsection{Fock sector dependent renormalization}\label{sect:FSDR}

Prior to solving the master equation (\ref{eqn:Gamma_chi_varphi_cc}), 
we need to know the bare parameters. According to Fock sector dependent renormalization \cite{Karmanov:2008br}, $g_\pi$, $g_{ N\bar N}$, $\delta \mu^2_{N\bar N}$ and $\delta m^2_\pi$ are determined from the two-body truncations, similar to the recursive Forest Theorem in perturbation theory. These results are collected in the Appendix. 
The mass counterterm $\delta m^2_{\pi\pi}$ only appears in the one-body equation -- we do not need it in Eq.~ (\ref{eqn:Gamma_chi_varphi_cc}). The bare coupling $g_{\pi\pi}$, associated with an interaction vertex that can be dressed up to two pions, is the only bare parameter that needs to be determined. We determine this parameter from the coupling constant renormalization.  
Up to the three-body truncation, the two-body vertex $\Gamma_{\pi N}$ can be decomposed
as,
\begin{equation}
\Gamma_{\pi N}(s) = \sqrt{Z_{2}} V_{\pi N N}(s) Z_{23}(s)
\end{equation}
where $Z_2$ is the on-shell $Z$-factor associated with the nucleon and $Z_{23}(s)$ is the off-shell $Z$ factor defined in Eq.~(\ref{eqn:hybridZ}), and 
$V_{\pi N N}$ is the amputated 3-point vertex. All off-shell quantities depends on the kinematical variable $s = (k^2_\perp+\mu^2)/x+(k^2_\perp+m^2)/(1-x)$, which is the invariant mass of the $\pi N$ Fock state.

The Lehmann-Symanzik-Zimmermann reduction formula relates $V_{\pi N N}$ to the physical coupling $g$, defined as the tree-level on-shell scattering amplitude $T_{fi}^\star$, to the elementary vertex at the mass shell point $s^\star = m^2$ \cite{Lehmann:1954rq, Lehmann:1957zz},
\begin{equation}
\sqrt{Z_2} V_{\pi N N} (s^\star = m^2)\sqrt{Z_2}\sqrt{Z_3} = T_{fi}^\star = g.
 \end{equation}
Therefore, the suitable renormalization condition for our system within three-body truncation should be, 
\begin{equation}
\Gamma_{\pi N}(x^\star, k^\star_\perp) = g \frac{Z_{23}}{\sqrt{Z_2Z_3}}, \quad \Big( s^\star = 
\frac{k^{\star2}_\perp+m^2}{1-x} + \frac{k^{\star2}_\perp+\mu^2}{x}  = m^2 \Big).
\end{equation}
Imposing this condition to Eq.~(\ref{eqn:Gamma_chi_varphi_cc}), we get,
\begin{multline}\label{eqn:Gamma_chi_varphi_ren}
\frac{g}{\sqrt{Z_\pi Z_{ N\bar N}}} = g_{\pi\pi}\psi_ N  
  +\  g_\pi^2 \int\limits_0^{1-x} \frac{\dd x'}{2x'(1-x')(1-x-x')}
\int\frac{\dd^2k'_\perp}{(2\pi)^3} \frac{\Gamma_{\pi N}(x', k'_\perp)}{s'_{2}-M^2} 
\frac{1}{s_{3c}^\star-M^2}  \\
+ g_{ N\bar N}^2 \int\limits_{1-x}^1 \frac{\dd x'}{2x'(1-x')(x+x'-1)}\int\frac{\dd^2k'_\perp}{(2\pi)^3} \frac{\Gamma_{\pi N}(x', k'_\perp)}{s'_{2}-M^2} 
\frac{1}{s_{3d}^\star-M^2}. 
\end{multline}
This equation fixes the bare coupling $g_{\pi\pi}$ in terms of the physical coupling as well as the vertex function $\Gamma_{\pi N}$, which is yet to be solved. 
Combining Eq.~(\ref{eqn:Gamma_chi_varphi_cc}) with Eq.~(\ref{eqn:Gamma_chi_varphi_ren}), we can exclude $g_{\pi\pi}\psi_N$ and obtain:
\begin{multline}\label{eqn:Gamma_chi_varphi_dd}
 \bigg[ 
 1 -
 \frac{\Sigma_\pi^R(t_a)}{t_a - m^2}   - 
\frac{\Pi_{ N\bar N}^R(t_b)}{t_b-\mu^2}  \bigg]
\Gamma_{\pi N}(x, k_\perp)
= \frac{g}{\sqrt{Z_\pi Z_{ N\bar N}}} \\
  +  g_\pi^2 \int\limits_0^{1-x} \frac{\dd x'}{2x'(1-x')(1-x-x')}
\int\frac{\dd^2k'_\perp}{(2\pi)^3} \frac{\Gamma_{\pi N}(x', k'_\perp)}{s'_{2}-M^2} 
\Big( \frac{1}{s_{3c}-M^2} - \frac{1}{s_{3c}^\star-M^2} \Big)  \\
+ g_{ N\bar N}^2 \int\limits_{1-x}^1 \frac{\dd x'}{2x'(1-x')(x+x'-1)}\int\frac{\dd^2k'_\perp}{(2\pi)^3} \frac{\Gamma_{\pi N}(x', k'_\perp)}{s'_{2}-M^2} 
\Big( \frac{1}{s_{3d}-M^2} - \frac{1}{s_{3d}^\star-M^2} \Big), 
\end{multline}
where,
\begin{align}
s'_2 =\,& \frac{k'^2_\perp+\mu^2}{x'} + \frac{k'^2_\perp+m^2}{1-x'}, \\
s_{3c}=\,&  \frac{k^2_\perp+\mu^2}{x} + \frac{k'^2_\perp+\mu^2}{x'} + \frac{(\vec k_\perp+\vec k'_\perp)^2+m^2}{1-x-x'}, \\
s_{3d}=\,&   \frac{k^2_\perp+m^2}{1-x} + \frac{k'^2_\perp+m^2}{1-x'} + \frac{(\vec k_\perp+\vec k'_\perp)^2+m^2}{x+x'-1}, \\
t_a =\,& m^2 - (1-x) (s_{2} - M^2), \\
t_b =\,& \mu^2 - x(s_{2} - M^2), \\
s_{2} = \, & \frac{k^2_\perp+\mu^2}{x} + \frac{k^2_\perp+m^2}{1-x}
.
\end{align}
As mentioned, the bare parameters $g_\pi$, $g_{N\bar N}$ as well as the self-energies $\Sigma^R_{\pi}(t)$ and $\Pi^R_{N\bar N}(t)$ are obtained in the two-body truncation as shown in the Appendix. The kinematical domains of the off-shell self-energies are
$t_a \in \big(-\infty, (m-\mu)^2\big]$, and $t_b \in \big(-\infty, 0\big]$.  In these domains, there are no singularities or branch cuts in the 
self-energies.

\subsection{Numerical solutions}

\begin{figure} \centering
\includegraphics[width=0.55\textwidth]{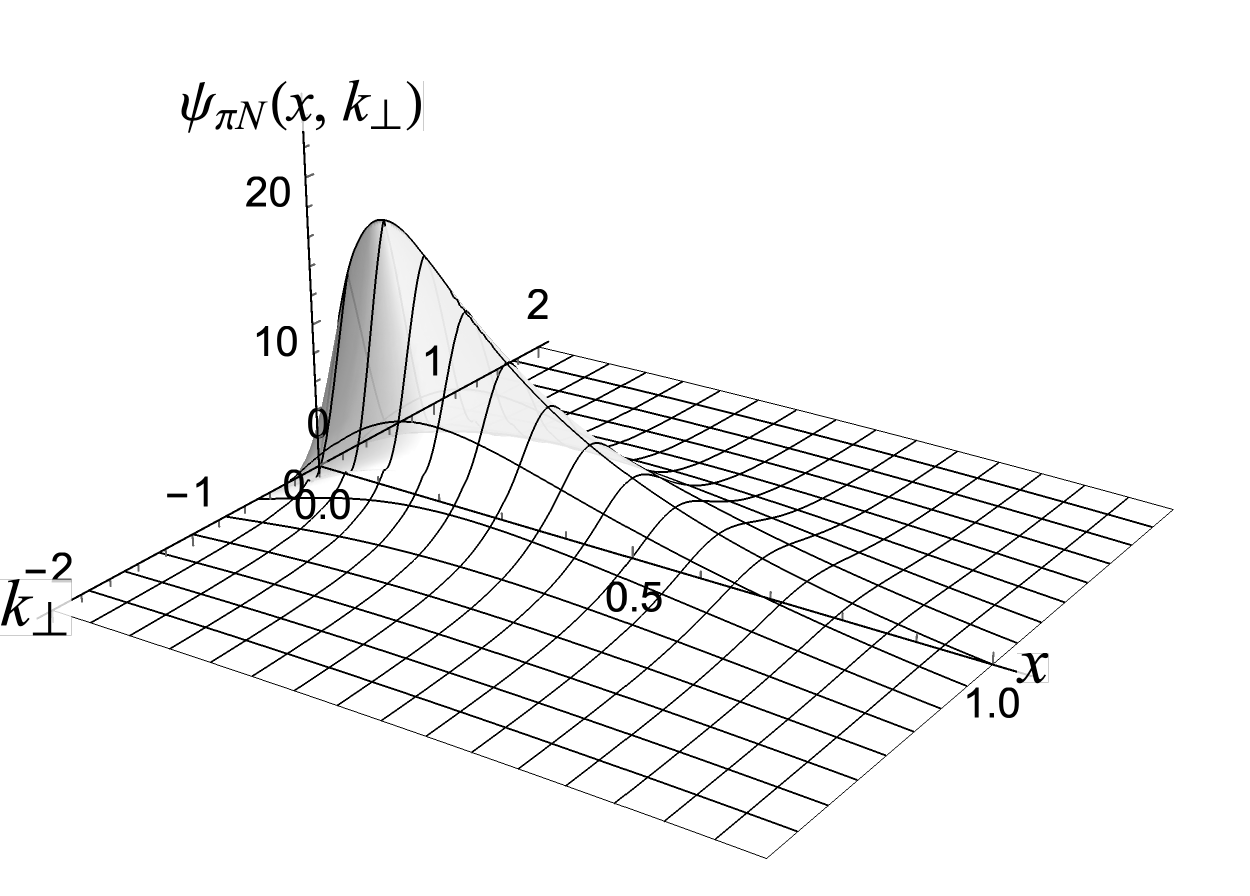}
\caption{The two-body wave function $\psi_{\pi N}(x, k_\perp)$ solved from the system of equations (\ref{eqn:Gamma_chi_varphi_dd}) with $\alpha = 1.0$. }
\label{fig:wf}
\end{figure}

Equation~(\ref{eqn:Gamma_chi_varphi_dd}) is the main equation to solve for the wave functions. We discretize the variables $x, k_\perp, \theta_{k}$ using a Lagrange mesh, i.e. Legendre quadratures. The radial is first mapped to $(0, 1)$ from $(0, \infty)$ before applying the quadrature mesh. Then, the integrals can be approximated using Gauss quadrature. To solve the integral equation (\ref{eqn:Gamma_chi_varphi_dd}), we designed an iterative procedure. The numerical tolerance is set to,
\begin{equation}
\sup \big|\Gamma_{\pi N}^{(n)} - \Gamma_{\pi N}^{(n-1)}\big| \le 10^{-5}.
\end{equation}
The converged vertex $\Gamma_{\pi N}$ then can be used to evaluate observables. The wave functions are related to the vertex functions by a simple energy denominator, 
\begin{equation}
\psi_n(\{x_i, k_{i\perp}\}) = \frac{\Gamma_n(\{x_i, k_{i\perp}\})}{s_n - M^2},
\end{equation}
where, 
\begin{equation}
s_n = \sum_i \frac{k^2_{i\perp}+m_i^2}{x_i}
\end{equation}
is the $n$-body invariant mass squared.  
Fig.~\ref{fig:wf} shows the obtained wave function $\psi_{\pi N}(x, k_\perp)$ at $\alpha = 1.0$. 

\begin{figure} \centering
\includegraphics[width=0.6\textwidth]{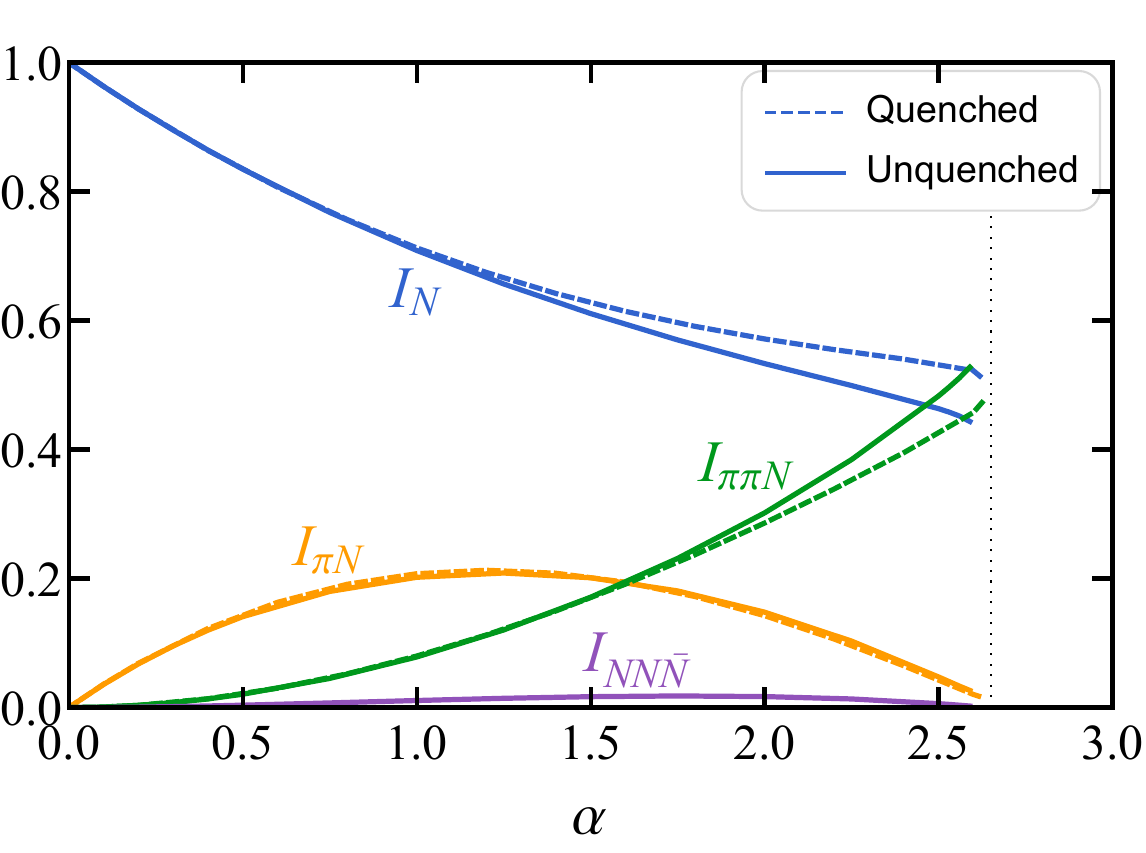}
\caption{(Colors online) The normalization of the Fock sectors $I_N$, $I_{\pi{}N}$, $I_{\pi\pi{}N}$ and $I_{NN\bar{N}}$ as a function of the coupling $\alpha$. The solid curves are the full (unquenched) three-body truncation results while the dashed curves are our previous three-body truncation results with a quenched approximation. The dotted vertical line indicates the critical coupling $\alpha_c = 2.63$ for the Landau pole.
}
\label{fig:In}
\end{figure}

One set of quantities of interest consists of the normalization of each Fock sector, also known as the Fock components, 
\begin{equation}
I_n = \frac{1}{S_n}\prod_{i} \int \frac{\dd x_i}{2x_i} \int \frac{\dd^2 k_{i\perp}}{(2\pi)^3} 2\delta(\sum_i x_i - 1)(2\pi)^3\delta^2(\sum_i \vec k_{i\perp}) 
\Big|\psi_n(\{x_i, \vec k_{i\perp}\}) \Big|^2.
\end{equation}
Here, $S_n$ is the symmetry factor, for example, $S_{\pi\pi N} = 2!$. 
Adding up all Fock components gives one,
\begin{equation}
\sum_n I_n = 1.
\end{equation}
In our case, the three-body truncation, there are 4 Fock sectors, with corresponding  $I_N$, $I_{\pi N}$, $I_{\pi\pi N}$ and $I_{NN\bar N}$. 
Figure~\ref{fig:In} shows the normalization of each Fock sector as a function of the coupling $\alpha$ up to the critical coupling $\alpha_c = 2.63$ for the Landau pole \cite{Karmanov:2016yzu}.  Our previous results from a three-body truncation but with a quenched approximation are also shown for comparison \cite{Karmanov:2016yzu}. 
From these results, the $| N N\bar N\rangle$ Fock sector takes only a very small portion of the state vector ($\lesssim 2\%$), which validates our previous quenched approximation. However, as we will show in Sect.~\ref{sect:EMFF}, this Fock sector plays an important role in preserving the covariance of the e.m. form factor.  

The normalization of the single nucleon sector $I_N$ is of particular interests. This quantity is precisely the on-shell $Z_2$ factor for the field strength normalization. The finiteness of $I_N = Z_2$ suggests that the physical nucleon has a pointlike core. 

\section{Covariant analysis of the form factor}\label{sect:EMFF}

The nucleon field $ N$ is charged and can couple to the photon by minimal coupling: $\partial_\mu \to D_\mu = \partial_\mu + i e A_\mu$. 
The corresponding current operator is:
\begin{equation}
J^\mu = i (D^\mu  N)^\dagger  N - i  N^\dagger D^\mu  N = i \partial^\mu  N^\dagger  N - i  N^\dagger \partial^\mu  N
+ 2 e A^\mu  N^\dagger  N.
\end{equation}
With this operator defined and the wave functions for the physical state available, we are now ready to compute the HME $\langle p' |J^\mu(0) | p\rangle$ for the scalar Yukawa theory. 

\begin{figure}
\centering
\includegraphics[width=.7\textwidth]{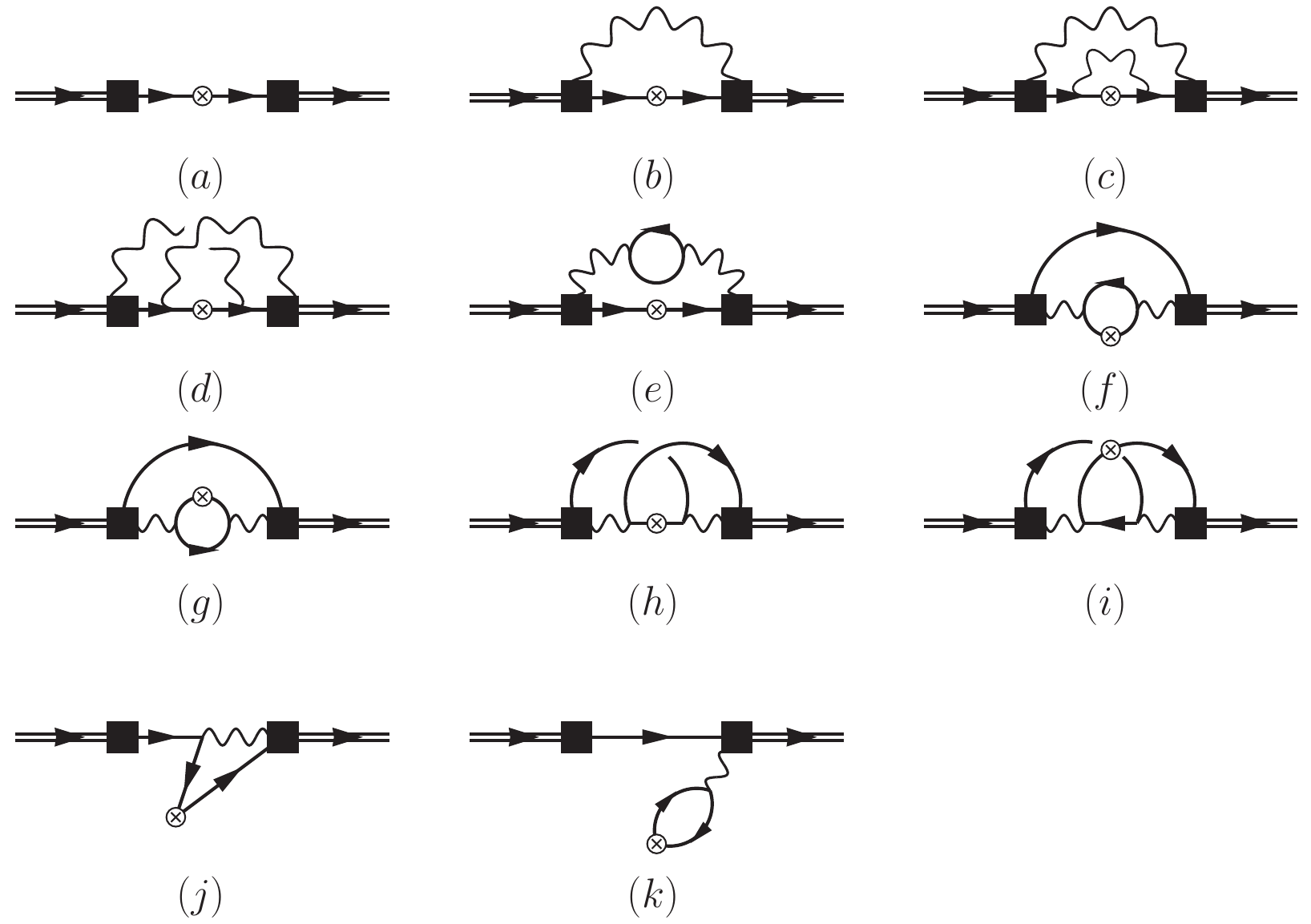} 
\caption{Diagrams contributed to the hadronic matrix element of the e.m. current operator within the three-body truncation. The double lines represent the physical nucleon. The solid lines represent the nucleon and the wavy lines the pion. The crossed circles are the current operator insertion. The shaded boxes are the vertex functions, which is directly related to the wave functions. We have made use of the equation of motion to replace the three-body vertex functions with two-body vertex functions. Note that for diagrams ($j$)--($k$), their Hermitian conjugates (not shown) also contribute. 
}
\label{fig:diagrammatics}
\end{figure}

Figure~\ref{fig:diagrammatics} shows the diagrammatic representation of the current HME within the three-body truncation. 
The double lines represent the physical nucleon. The solid lines represent the nucleon and the wavy lines the pion. The crossed circles are the current operator insertion. The shaded boxes are the vertex functions that we obtained in Sect.~\ref{sect:threey-body_truncation}. These diagrams can be evaluated using the light-front Feynman rules. 
Note that we neglect the seagull term $2 e A^\mu  N^\dagger  N$, since it is in the next-to-leading order in $\alpha_\text{em} = e^2/4\pi$.
We have used the equation of motion to replace the three-body vertex functions by the two-body vertex function $\Gamma_{\pi N}$ (shaded boxes). The effect of the higher Fock sectors are the dressing of the current, similar to Dyson-Schwinger equations. 

Diagrams ($a$)--($i$) are all diagonal contributions, i.e. they involve the overlap of wave functions of the same number of particles. Diagrams ($j$)--($k$) involves the pair creation term, also known as the $Z$-term. Even though they are not diagonal, they still involve the overlap of wave functions, except with different numbers of particles.

\subsection{Covariant light front dynamics} 

By virtue of the Poincaré symmetry, the HME of the current operator can be written as, 
\begin{equation}
\langle p' | J^\mu(0) | p \rangle = (p+p')^\mu F(-q^2)
\end{equation}
where, $q = p' - p$, $Q^2 = -q^2$ and $F(Q^2)$ is known as the e.m. form factor. Physically, it represents the Fourier transform of the charge density. This interpretation is only exact on the light front in the Drell-Yan frame. 
The above Lorentz decomposition assumes the full Poincaré symmetry. As we have seen, in practical calculations like in this work, the full Poincaré symmetry may be violated by truncations including the Fock sector truncation. As such, the symmetry within the wave function is reduced to a subgroup of the Poincaré symmetry and we should parametrize the HME according to this reduced symmetry. 
New covariant structures emerge with the reduced symmetry, which vanish if full Poincaré symmetry is restored in the continuum limit. For this reason, form factors associated with the new structures are called spurious form factors. 

Karmanov et. al. proposed a simple way to construct the new structures \cite{Karmanov:1991fv, Karmanov:1994ck, Karmanov:1996un, Carbonell:1998rj, Carbonell:1999pt, Brodsky:2003pw}. The covariant structures are constructed by adding $\omega^\mu$ to the set of relevant covariant tensors, e.g. $P^\mu, q^\mu$. 
In this approach, known as the covariant light-front dynamics (CLFD), the covariant decomposition of the HME of the e.m. current operator is,
\begin{equation}\label{eqn:current_clfd}
j^\mu \equiv \langle p'|J^\mu(0)|p\rangle = (p+p')^\mu F(\zeta, Q^2) + \frac{\omega^\mu M^2}{\omega\cdot P} S_1(\zeta, Q^2)  + \frac{q^{[\mu} \omega^{\nu]}q_\nu}{\omega\cdot P} S_2(\zeta, Q^2),
\end{equation}
where $q = p'-p$, $P = (p'+p)/2$, $Q^2 = -q^2$, $a^{[\mu}b^{\nu]} = a^\mu b^\nu - a^\nu b^\mu$ and $\zeta = \omega\cdot q/\omega \cdot P = q^+/P^+$. Note that the form factors depend on two scalars, $\zeta$ and $Q^2$. 
$F$ is the physical form factor, and  $S_{1,2}$ are spurious form factors due to the possible violation of the Poincaré symmetry. The spurious structure associated with $S_1$  also violates current conservation. One expects that both spurious form factors vanish in the continuum limit.
Hermiticity of the current operator requires $F(\zeta, Q^2) = F^*(-\zeta, Q^2)$ and $S_{1,2}(\zeta, Q^2) = S^*_{1,2}(-\zeta, Q^2)$. 

In previous work in the literature, the Drell-Yan frame $\zeta = 0$ is taken. There are several dramatic simplifications in the Drell-Yan frame.  First, Hermiticity implies that form factors in the Drell-Yan frame are real functions. Second, the spurious structure $\propto q^\mu$ vanishes.  Furthermore, the e.m. current is conserved in the Drell-Yan frame. 
By Eq.~(\ref{eqn:current_clfd}), we extend the CLFD analysis to the most general frames. In a general frame, the Lorentz structure ${M^2\omega^\mu }/{\omega\cdot P}$ does not conserve the current. Hence we do not assume current conservation a priori. To take advantage of the boost invariance in light-front dynamics, we introduce a boost-invariant transverse momentum, $\vec \Delta_\perp = \vec q_\perp - \zeta \vec P_\perp$. We now have two external boost invariants $\zeta$ and $\vec\Delta_\perp$ following Eq.~(\ref{eqn:boosts}), which are related to the external Lorentz invariant $Q^2$ as, 
\begin{equation}
Q^2 = \frac{\zeta^2M^2 + \Delta_\perp^2}{1 - \frac{1}{4}\zeta^2}.
\end{equation}
The dependence of the form factor on $\zeta$ for fixed $Q^2$ is the frame dependence within the form factor, i.e. each $\zeta$ defines a class of reference frames that are related by light-front boosts. 
In Ref.~\cite{Li:2017uug}, we introduced two special frames: 
\begin{enumerate}  
\item Transverse frame (Drell-Yan frame): $\zeta = 0$; 
\item Longitudinal frame: $\Delta_\perp = 0$
\end{enumerate}
We showed the spread of the form factors that lies within these two frames. Note that both $\zeta$ and $\vec \Delta_\perp$ are invariant under the longitudinal and transverse light-front boosts. As such, each frame represents a large class of frames that are related by boosts. In the literature, e.g. Ref.~\cite{Sawicki:1992qj}, the longitudinal frame is defined as $\vec q_\perp = 0$, which is not boost invariant. Our definition agrees with this definition only at $\vec P_\perp = 0$. 

To extract the physical form factor $F$, we consider the current matrix elements, 
\begin{align}
j^+ =\,& 2P^+ F(\zeta, Q^2) + \zeta^2 P^+S_2(\zeta, Q^2), \\
\vec j_\perp =\,& 2\vec P_\perp F(\zeta, Q^2) + \zeta \vec q_\perp S_2(\zeta, Q^2).
\end{align}
Then, the physical form factor are, 
\begin{equation}
F(\zeta, Q^2) = \frac{j^+}{2P^+} \frac{\vec q_\perp}{\vec \Delta_\perp} - \zeta \frac{\vec j_\perp}{2\vec\Delta_\perp}. 
\end{equation}
These expressions can be further simplified in the transverse Breit frame\footnote{It is also known as the symmetric frame.} $\vec P_\perp = 0$, 
\begin{align}
j^+ =\,& 2P^+ F(\zeta, Q^2) + \zeta^2 P^+S_2(\zeta, Q^2), \\
\vec j_\perp =\,& \zeta \vec q_\perp S_2(\zeta, Q^2).
\end{align}
The physical form factor can be extracted as, 
\begin{equation}
F(\zeta, Q^2) = \frac{j^+}{2P^+} - \zeta \frac{\vec j_\perp}{2\vec q_\perp}.
\end{equation}
In the Drell-Yan frame ($\zeta = 0$), this expression reduces to the familiar $J^+$ current, 
\begin{equation}
F(0, Q^2) = \frac{j^+}{2P^+}.
\end{equation}

\subsection{Zero-mode contributions} \label{sect:zero_modes}

One of the important questions is whether the $Z$-terms contain zero modes in the Drell-Yan frame $q^+ = 0$, which concerns diagrams ($j$)--($k$) in Fig.~\ref{fig:diagrammatics}. If there are no zero-mode contributions, these diagrams should vanish in the Drell-Yan frame $q^+ = 0$, leaving only the diagonal contributions to the form factor, as described by the Drell-Yan-West formula.  To detect the possible zero-mode contributions, we evaluate these terms in a frame without requiring $q^+=0$, and take the limit $q^+ \to 0$ at the end. 

In a general frame, the contribution to the form factor from diagram ($j$) is, 
\begin{multline}
F_j(\zeta, Q^2) = g_{N\bar N}\sqrt{Z_2} (2+\zeta)^2 \int_0^\zeta \frac{\dd z}{2z(\zeta - z)(2+\zeta-2z)} \int \frac{\dd^2 l_\perp}{(2\pi)^3} \\\times \psi_{\pi N}\Big(1 - \frac{2z}{2+\zeta}, \vec l_\perp - \frac{z}{2+\zeta}\vec q_\perp\Big) \frac{z - \zeta \frac{\vec l_\perp}{\vec q_\perp}}{s_3 - m^2}
\end{multline} 
where, recall $\zeta = q^+/P^+$,  and 
\begin{equation}
s_3 = \frac{\big(\vec l_\perp - \frac{z}{2+\zeta}\vec q_\perp\big)^2 + m^2}{\frac{2z}{2+\zeta}} + \frac{\big(\frac{2}{2+\zeta}q_\perp\big)^2+m^2}{\frac{2-\zeta}{2+\zeta}} + \frac{\big(\vec l_\perp - \frac{z+2}{2+\zeta}\vec q_\perp\big)^2 + m^2}{1 - \frac{2z}{2+\zeta} - \frac{2-\zeta}{2+\zeta}}.
\end{equation}
We have adopted the Breit frame $\vec P_\perp = 0$ without the loss of generality. Let $x \equiv z/\zeta$, $a \equiv 2\zeta/(2+\zeta)$, $\vec k_\perp = \vec l_\perp - x\vec q_\perp$. In the Drell-Yan limit $\zeta \to 0$, $F_j$ becomes, 
\begin{multline}
\lim_{\zeta \to 0} F_j(\zeta, Q^2) = \lim_{a \to 0} g_{N\bar N}\sqrt{Z_2} \frac{a^2(1-a)}{a-2} \int_0^1 x \dd x \int \frac{\dd^2 k_\perp}{(2\pi)^3} \frac{\vec k_\perp}{\vec q_\perp} \\
\times \frac{\Gamma_{\pi N}\Big(1 - ax, \vec k_\perp + (1-\frac{1}{2}a)x\vec q_\perp\Big)}{\big(\vec k_\perp+(1-\frac{1}{2}a)x\vec q_\perp\big)^2 + a x\mu^2 + (1-ax)^2 m^2} \\
\times \frac{1}{(1-a)(k_\perp^2+m^2)+x(1-x)\big((1-\frac{1}{2}a)^2 q_\perp^2+a^2 m^2\big)}.
\end{multline} 
Observe that the integral now is finite\footnote{Note that $\Gamma_{\pi N}$ is finite at the endpoints} as $a$ approaches 0. Therefore, 
\begin{equation}
\lim_{\zeta \to 0} F_j(\zeta, Q^2) = 0.
\end{equation}

Similarly, the contribution from diagram ($k$) is,
\begin{multline}
F_k(\zeta, Q^2) = g_{N\bar N}\sqrt{Z_2} \frac{(2+\zeta)^2}{4\zeta} \psi_{\pi N}\Big(\frac{2\zeta}{2+\zeta}, -\frac{2}{2+\zeta}\vec q_\perp\Big) \\
\times \int_0^\zeta \frac{\dd z}{2z(\zeta - z)} \int \frac{\dd^2l_\perp}{(2\pi)^3} \frac{z - \zeta \frac{\vec l_\perp}{\vec q_\perp}}{s_3 - m^2}
\end{multline} 
Let $x \equiv z/\zeta$, $a \equiv 2\zeta/(2+\zeta)$, $\vec k_\perp = \vec l_\perp - x\vec q_\perp$. In the Drell-Yan limit $\zeta \to 0$, $F_k$ becomes, 
\begin{multline}
\lim_{\zeta \to 0} F_k(\zeta, Q^2) = \lim_{a \to 0} g_{N\bar N}\sqrt{Z_2}  \frac{1-a}{a-2} \psi_{\pi N}\big(a, -(1-\frac{1}{2})\vec q_\perp\big) \\
\times \int_0^1 \dd x \int \frac{\dd^2k_\perp}{(2\pi)^3}\frac{\vec k_\perp}{\vec q_\perp} \frac{1}{(1-a)(k_\perp^2+m^2) + x(1-x)\big((1-\frac{1}{2}a)^2q_\perp^2 + a^2 m^2\big)}.
\end{multline} 
The above expression vanishes for two reasons. First, the transverse integral involves an odd integrand caused by $\vec k_\perp$. Second, the wave function $\psi_{\pi N}$ vanishes at the endpoint $a \to 0$. In any case, we have, 
\begin{equation}
\lim_{\zeta \to 0} F_k(\zeta, Q^2) = 0.
\end{equation}
Therefore, the pair creation/annihilation terms vanish in the limit $q^+ \to 0$, and there is no zero-mode contributions to the e.m. current in the Drell-Yan frame for the scalar Yukawa theory.  This conclusion is consistent with the observation by Brodsky~et.~al. based on LCPT \cite{Brodsky:1998hn}. 
Some authors, e.g. Refs.~\cite{Choi:1998nf, Choi:2021xld}, argued that $J^-$ is associated with the zero modes. From the covariant analysis, this current $J^-$ is associated with the spurious form factor $S_1$. Therefore, it should be used with caution.

\subsection{Frame dependence of the form factor}\label{sect:frame_dependence}

\begin{figure} \centering
\subfigure[\ \label{fig:FF_two-body}]{\includegraphics[width=0.6\textwidth]{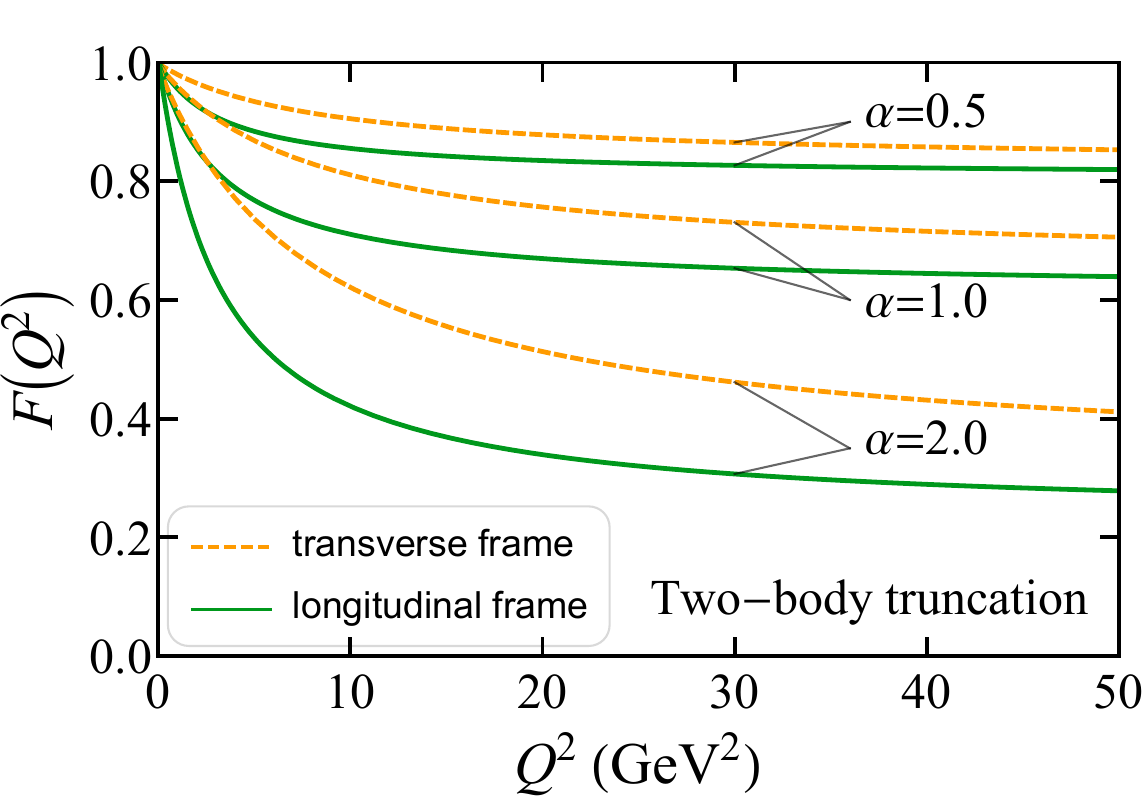}} 
\subfigure[\ \label{fig:FF_three-body}]{\includegraphics[width=0.6\textwidth]{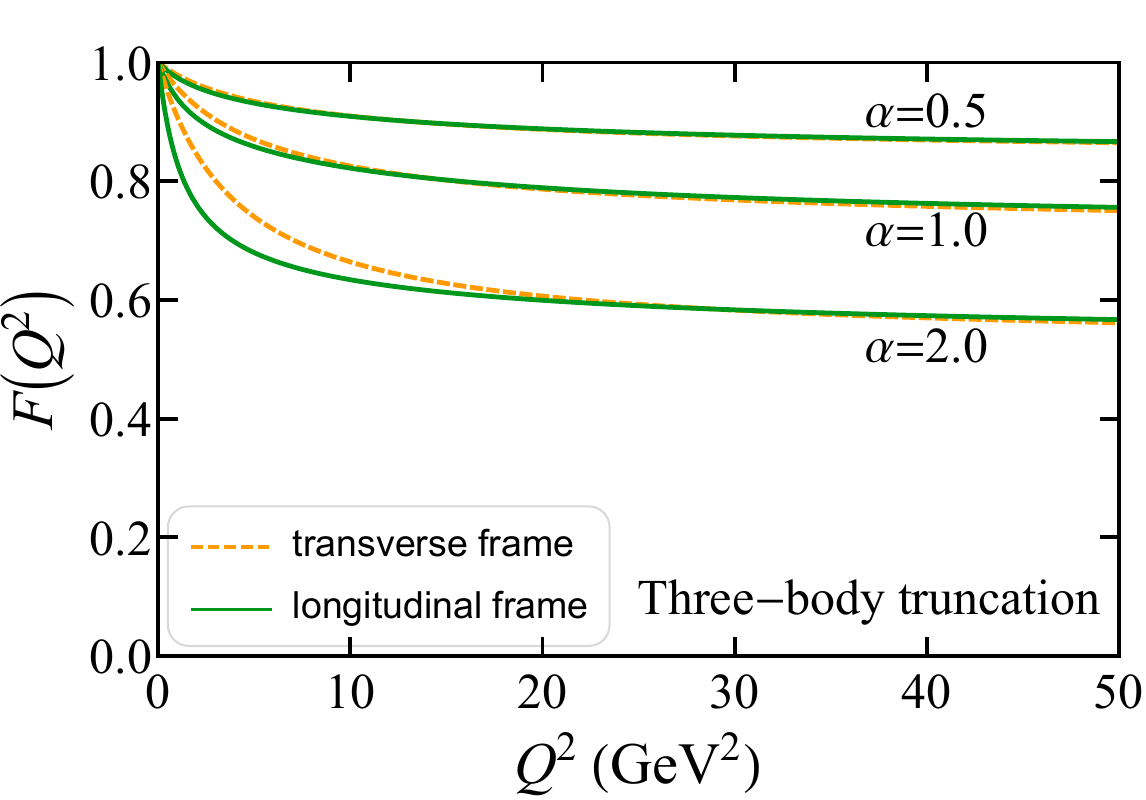}}
\caption{Comparison of the e.m. form factors obtained in the transverse (Drell-Yan) frame and the longitudinal frame at selected couplings.
(\textit{Left}) Results from two-body truncation ($|N\rangle + |\pi N\rangle$); (\textit{Right}) Results from three-body truncation ($|N\rangle + |\pi N\rangle + |\pi \pi N\rangle + |N N \bar N\rangle$).}
\label{fig:FF_frame_dependence}
\end{figure}

In this subsection, we investigate the frame dependence of the form factor. 
Figure~\ref{fig:FF_frame_dependence} compares form factors evaluated in two selected frames, the Drell-Yan frame (transverse frame) and the longitudinal frame, with a two-body Fock sector truncation ($|N\rangle_\text{ph} = |N\rangle + |\pi N\rangle$, Fig.~\ref{fig:FF_two-body}) and a three-body Fock sector truncation ($|N\rangle_\text{ph} = |N\rangle + |\pi N\rangle + |\pi \pi N\rangle + |N N \bar N\rangle$, Fig.~\ref{fig:FF_three-body}). In each case, the wave functions are solved within the corresponding Fock space first. Wave functions for two-body truncation are solved previously in Refs.~\cite{Li:2014kfa, Li:2015iaw, Karmanov:2016yzu} (see also the Appendix). Wave functions for the three-body truncation are obtained in Sect.~\ref{sect:threey-body_truncation}. 
The frame dependence of the form factors, i.e. the discrepancy between results evaluated at these two frames, indicates the violation of the Poincaré symmetry. From these figures, form factors with two-body truncation, which is equivalent to leading-order LCPT, exhibit considerable frame dependence, even at relatively weak coupling $\alpha = 0.5$. On the other hand, the frame dependence is dramatically reduced as the three-body Fock sectors $ |\pi \pi N\rangle + |N N \bar N\rangle$ are included. Within the three-body truncation, the frame dependence is fairly small at small coupling and increases at the coupling $\alpha$ increases. The reduction of the frame dependence in the three-body truncation is also a sign of convergence of the Fock sector expansion. 

\begin{figure} \centering
\subfigure[\ \label{fig:FF_three-body_quenched}]{\includegraphics[width=0.6\textwidth]{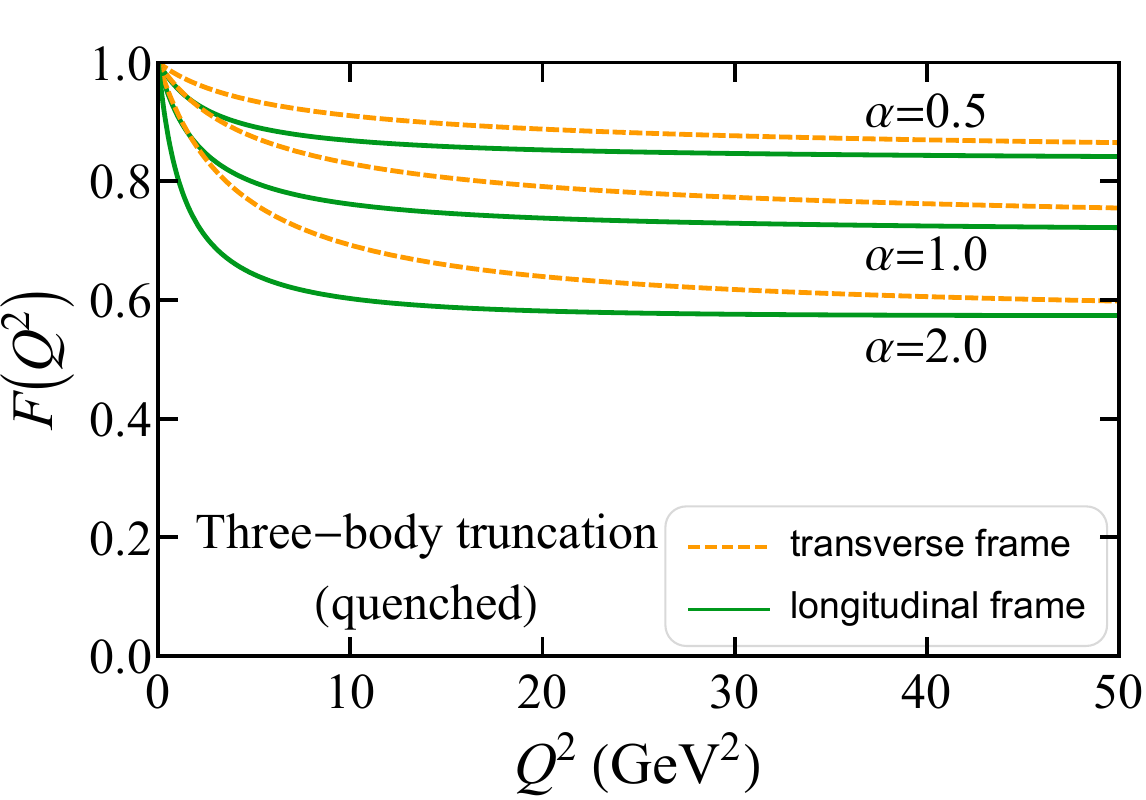}}
\subfigure[\ \label{fig:FF_three-body_unquenched}]{\includegraphics[width=0.6\textwidth]{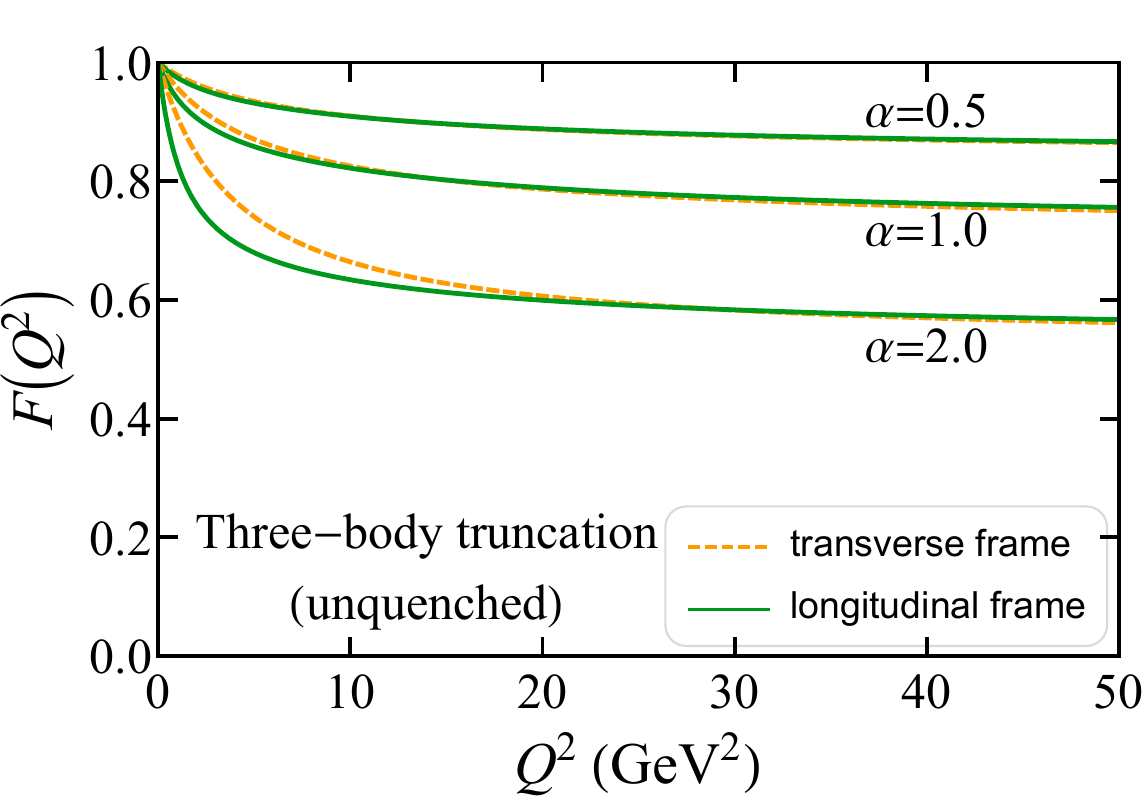}}
\caption{Comparison of the frame dependence of the e.m. form factors obtained with a quenched three-body truncation (\textit{Left})
and with  the unquenched three-body truncation (\textit{Right}) at selected couplings. The quenched three-body truncation does not include Fock sectors with anti-nucleon degrees of freedom, i.e. $|NN\bar N\rangle$. Even though this Fock sector contributes to a very small fraction ($\lesssim 2\%$) of the total probability of the state vector as shown in Fig.~\ref{fig:In}, the quenched results show a dramatically large frame dependence, as indicated by the discrepancy between form factors in the transverse frame and in the longitudinal frame. }
\label{fig:quenched_vs_unquenched_FF}
\end{figure}

To see the role of the pair creation/annihilation contribution ($Z$-term), we present the frame dependence of form factors evaluated within the quenched three-body truncation  ($|N\rangle_\text{ph} = |N\rangle + |\pi N\rangle + |\pi \pi N\rangle$, Fig.~\ref{fig:FF_three-body_quenched}), as well as within the full (unquenched) three-body truncation ($|N\rangle_\text{ph} = |N\rangle + |\pi N\rangle + |\pi \pi N\rangle+|NN\bar N\rangle$, Fig.~\ref{fig:FF_three-body_unquenched}). In the quenched case, the anti-nucleon degrees of freedom are excluded from the Fock space. Consequently,  the $Z$-term is absent in the e.m. current in Fig.~\ref{fig:diagrammatics}. 
While the Fock sector $|N N \bar N\rangle$ only contributes a very small portion ($\lesssim \,2\%$) of the total physical state (Fig.~\ref{fig:In}), 
the quenched results show a large frame dependence, as indicated by the discrepancy between form factors in the transverse and the  longitudinal frame, as shown by Fig.~\ref{fig:quenched_vs_unquenched_FF}.

\begin{figure} \centering
\subfigure[\ \label{fig:FF_three-body_Drell-Yan}]{\includegraphics[width=0.6\textwidth]{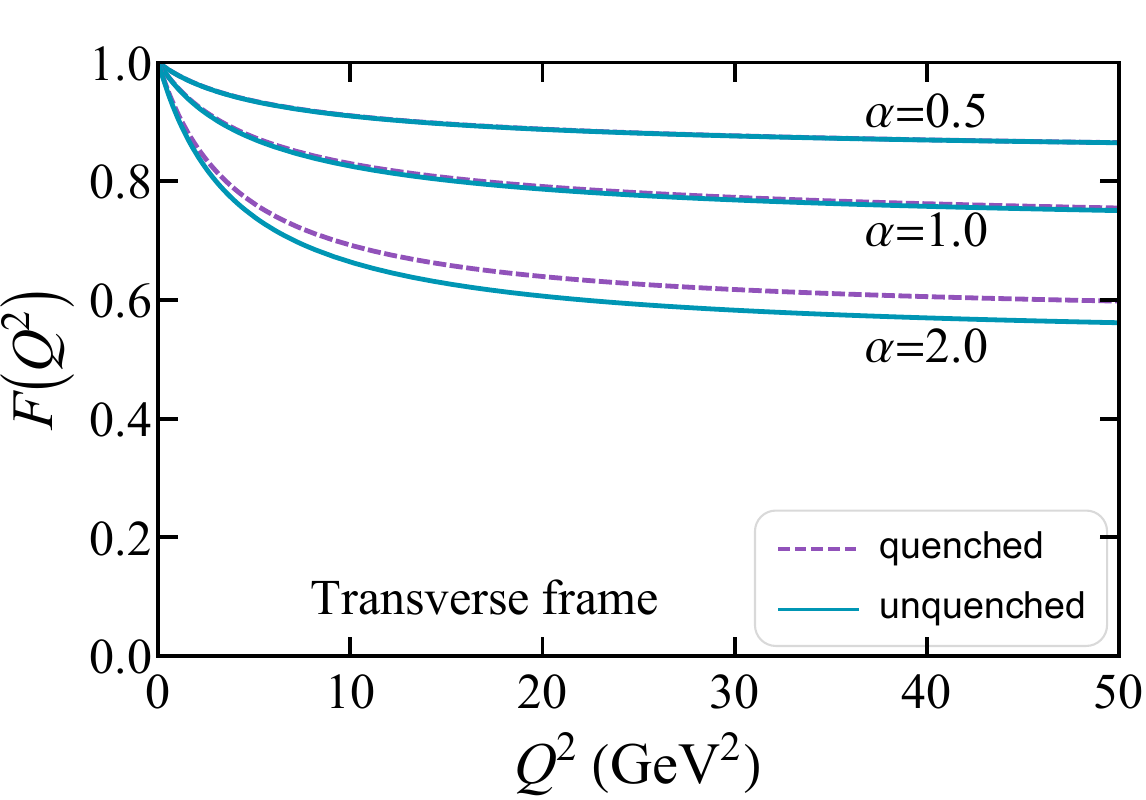}}
\subfigure[\ \label{fig:FF_three-body_longitudinal}]{\includegraphics[width=0.6\textwidth]{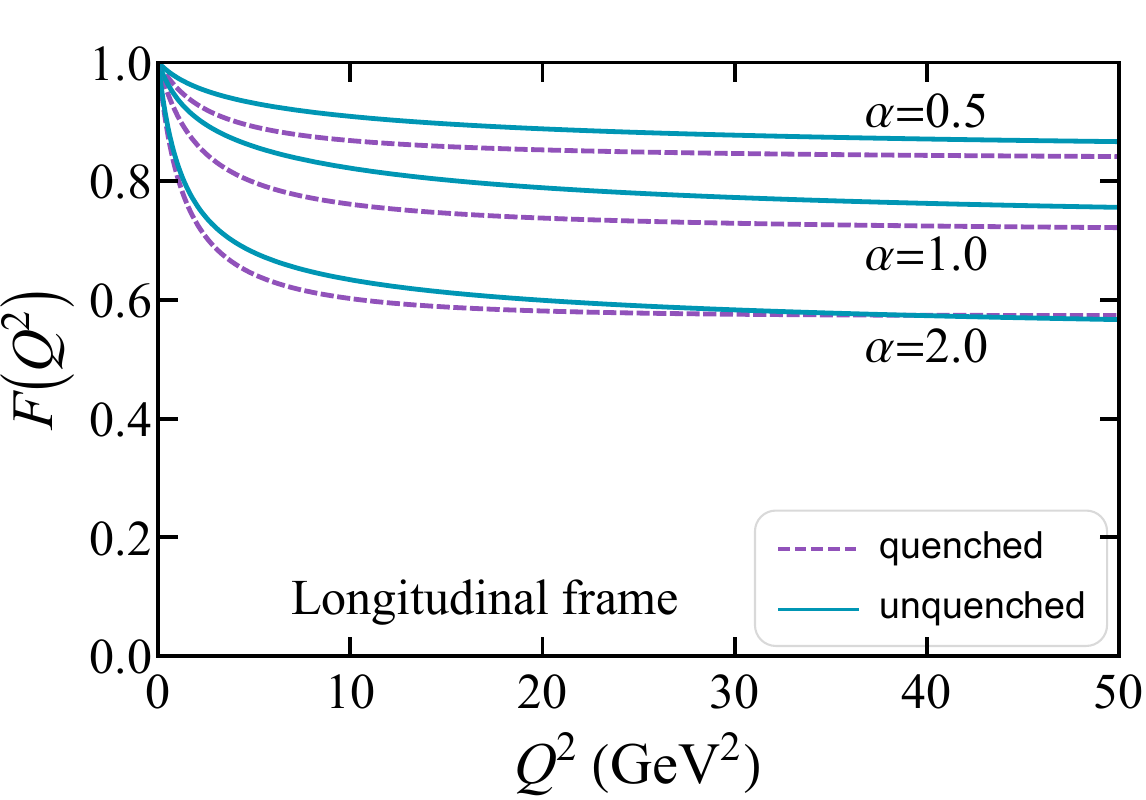}}
\caption{(\textit{Left}) Comparison of e.m. form factors from the quenched and unquenched three-body truncations in the transverse (Drell-Yan) frame at selected couplings;
(\textit{Right}) Comparison of e.m. form factors from the quenched and unquenched three-body truncations in the longitudinal frame at selected couplings.}
\label{fig:quenched_vs_unquenched_FF_frames}
\end{figure}

Figure~\ref{fig:quenched_vs_unquenched_FF_frames} compares the form factors obtained from the quenched and unquenched three-body truncations directly. Results within the transverse (Drell-Yan) frame are shown in Fig.~\ref{fig:FF_three-body_Drell-Yan}, and results within the longitudinal frame are shown in Fig.~\ref{fig:FF_three-body_longitudinal}. In the transverse frame, the form factors are quite close to each other, suggesting that the anti-nucleon degrees of freedom is not significant for form factors in this frame. Instead, the anti-nucleon degrees of freedom mainly impact form factors within the longitudinal frame. This conclusion corroborates with the analysis based on Bethe-Salpeter equations. On the other hand, the pair creation/annihilation contribution becomes accessible from wave functions -- see diagrams ($j$)--($k$) of Fig.~\ref{fig:diagrammatics}. The comparison of the quenched and the unquenched results also suggests that the Drell-Yan frame is a preferred frame within which the Fock sector expansion converges fast, in alignment with the conventional wisdom in light front physics \cite{Frankfurt:1977vc, Lepage:1980fj, Brodsky:1997de}.

\section{Summary and outlooks}\label{sect:summary}

In this work, we investigate the form factor of a strongly coupling scalar theory in (3+1)-dimensions within two different frames on the light cone: the transverse frame (Drell-Yan frame) and the longitudinal frame. The results are based on the non-perturbative solutions of the scalar Yukawa theory in full three-body truncation $|N\rangle + |\pi N\rangle + |\pi \pi N\rangle + | NN\bar N\rangle$ with a Fock sector dependent renormalization, which we also solve for the first time. By comparing results from two-body truncation as well as the quenched three-body truncation, we showed that the frame dependence, one of the effects of the violation of the Poincaré symmetry, is greatly reduced by incorporating the three-body Fock sectors. We also show the anti-nucleon degree of freedom plays an important role in restoring the Lorentz covariance. This is remarkable, since $| N N\bar N\rangle$ only populates a small portion of the state vector. 
From these results, we also find that the zero modes do not contribute to the current in the Drell-Yan frame for the scalar Yukawa theory. As a matter of fact, the form factor can be reliably extrapolated from the good current $J^+$ in the Drell-Yan frame, in agreement with the conventional wisdom. 

To the best of our knowledge, our work is the first systematic investigation of the Lorentz covariance of the form factor within light-front dynamics based on Fock sector expansion. Our results show that the Lorentz covariance as gauged by the frame dependence systematically improves as more Fock sectors are incorporated. With the help of covariant light-front dynamics, we show that a combination of $J^+$ and $\vec J_\perp$ can be used to extract the form factor in a general frame, without the contamination of the spurious contributions.

The scalar theory investigated in this work is relatively simple, and the Fock sector expansion employed in our work is straightforward. On the other hand, our work may serve as a cornerstone for developing more sophisticated methods for more complicated quantum field theories in the strong coupling regime. 
For example, the higher Fock sector contributions indicate how the e.m. current can be dressed on the light front. In Refs.~\cite{Duan:2024dhy, Cao:2024fto}, we also showed how to evaluate the energy-momentum tensor and parton distribution function beyond the valence sectors. 
With the increase of the computational capacity and the advance of theory, the light-front Hamiltonian formalism is becoming a promising tool to tackle strongly coupled quantum field theories. And the frame dependence can be used as a tool to improve the wave functions from these calculations. 

We investigate the zero-mode contributions in the current operator. However, there could also be zero modes in the state vector. Fortunately, for a massive scalar model, the zero modes are relatively simple -- they are non-dynamical and non-separated from the endpoints. For theories with dynamical zero modes, one should also incorporate their contributions. The massless scalar model is special. It is shown that for this theory, initial data at $x^+ =0$ is not sufficient to generate the full dynamics, and data on $x^- = -L$, where $L$ is an IR regulator, are also needed \cite{Heinzl:2018xnv}. These extra data can be interpreted as zero modes. In our current approach, we encountered various divergences when taking $\mu \to 0$ directly. It is interesting to investigate if the zero modes contained on $x^- = -L$ could help to resolve these problems. 

The present work adopts the Pauli-Villars regularization. This approach is non-perturbative, and is extendable to gauge theories \cite{Chabysheva:2015vga, Hiller:2015bic}. However, there are also known issues with this scheme. A few alternatives are available \cite{Pauli:1985pv, Pauli:1985ps, Vary:2009gt, Chabysheva:2011ed, Fitzpatrick:2018ttk}. See Ref.~\cite{Hiller:2016itl} for a partial review. Among these approaches,  Wilson and G\l{}azek proposed a Wilsonian approach in the light-front Hamiltonian formulation \cite{Glazek:1993rc, Glazek:1994qc}. In this approach, a low-energy-resolution effective Hamiltonian is first obtained from a similarity renormalization group evolution of the light-front QCD Hamiltonian. The effective Hamiltonian is then solved using standard quantum many-body methods \cite{Wilson:1994fk}. This Wilsonian approach is physical sound and computationally promising. However, the investigation of observables, including the frame-dependence of the current density, is more involved. Nevertheless, the method outlined in the present work is still applicable.

\section*{Acknowledgements}

We acknowledge fruitful discussions with Xingbo Zhao, Shaoyang Jia, Meijian Li, Wenyang Qian, Shuo Tang, Pieter Maris, and Vladimir A. Karmanov.
 This work was supported in part by the Chinese Academy of Sciences under Grant No.~YSBR-101, by the National Natural Science Foundation of China (NSFC) under Grant No.~12375081.

\appendix 

\section{Fock sector dependent renormalization of scalar Yukawa theory within two-body truncation}\label{sect:FSDR_two-body}

\subsection{Nucleon sector}

\begin{figure} \centering
\includegraphics[width=0.65\textwidth]{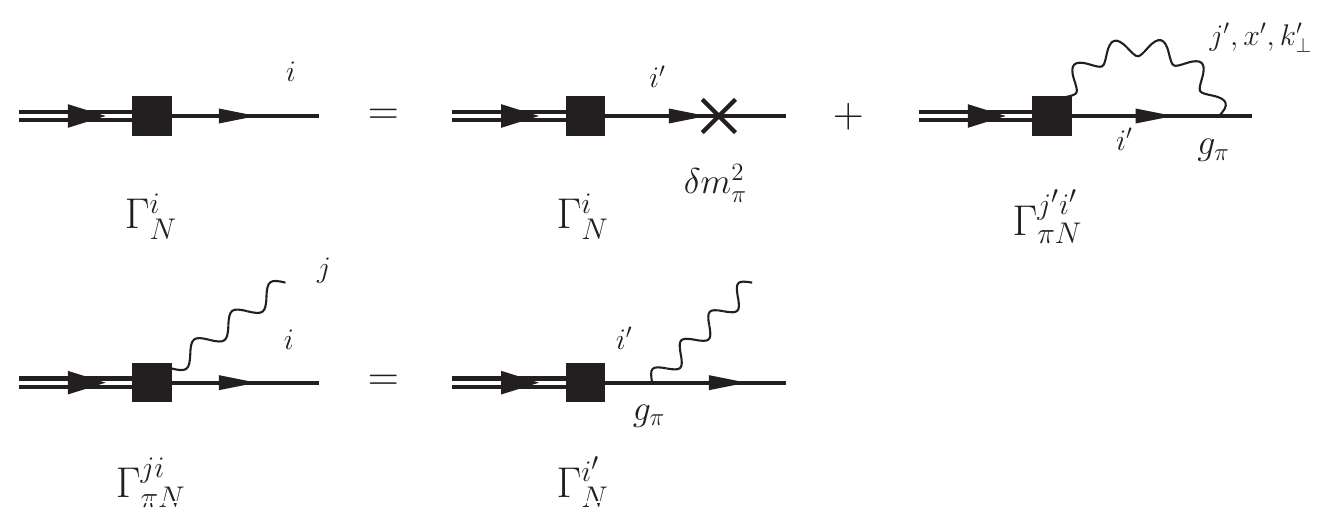}
\caption{Diagrammatic representation of the equation (\ref{eqn:Gamma_chi_varphi_N2}). The double lines represent the physical nucleon. The solid lines represent the nucleon or anti-nucleon. The wavy lines represent the pion. }
\label{fig:gamma_chi_varphi_N2}
\end{figure}

The renormalization parameters $\delta m_\pi^2$ and $g_\pi$ 
can be  found in Refs.~\cite{Li:2014kfa, Li:2015iaw, Karmanov:2016yzu}. Here we present 
the calculation for completeness. 
These parameters can be determined in the one-nucleon sector with a truncation up to $|N\rangle + |\pi N\rangle$, 
as shown in Fig.~\ref{fig:gamma_chi_varphi_N2}. From these diagrams, 
the system of equations is, 
\begin{align}\label{eqn:Gamma_chi_varphi_N2}
& \Gamma^i_ N = \delta m^2_\pi \sum_{i'}(-1)^{i'}\frac{\Gamma_ N^{i'}}{m^2_{i'}-M^2}
+ \sum_{i', j'} (-1)^{i'+j'}\int_0^1\frac{\dd x'}{2x'(1-x')}\int\frac{\dd^2k'_\perp}{(2\pi)^3}
\frac{g_\pi \Gamma_{\pi N}^{j'i'}(x', k'_\perp)}{s_2^{j'i'}-M^2}, \\
& \Gamma^{ji}_{\pi N}(x, k_\perp) = \sum_{i'}(-1)^{i'} g_\pi \frac{\Gamma_ N^{i'}}{m^2_{i'}-M^2}
\end{align}
where the energy denominator 
\begin{equation}
s_2^{j'i'} = \frac{k'^2_\perp+\mu^2_{j'}}{x'} + \frac{k'^2_\perp+m^2_{i'}}{1-x'}. 
\end{equation}

Renormalization condition: 
\begin{equation}
\Gamma_{\pi N}^{j=0,i=0}(x^\star, k^\star_\perp) = g, \quad 
\Big(\frac{k^{\star2}_\perp+\mu^2}{x^\star} + \frac{k^{\star2}_\perp+m^2}{1-x^\star} = m^2 \Big)
\end{equation}

We can remove the regulators, as shown for the master equation (\ref{eqn:Gamma_chi_varphi_cc}), 
and take the limit $M\to m$.  
\begin{align}\label{eqn:Gamma_chi_varphi_N2_b}
&0 = \delta m^2_\pi  \psi_ N
+  \int_0^1\frac{\dd x'}{2x'(1-x')}\int\frac{\dd^2k'_\perp}{(2\pi)^3}
\frac{g_\pi \Gamma_{\pi N}(x', k'_\perp)}{s_2-m^2}, \\
& \Gamma_{\pi N}(x, k_\perp) = g_\pi  \psi_ N, \\
& \Gamma_{\pi N}(x^\star, k^\star_\perp) = g \quad 
\Big(\frac{k^{\star2}_\perp+\mu^2}{x^\star} + \frac{k^{\star2}_\perp+m^2}{1-x^\star} = m^2 \Big)
\end{align}

From these relations, 
\begin{equation}
g_\pi = g \psi_ N^{-1}
\end{equation}

$\psi_ N$ can be obtained from the normalization condition:
\begin{equation}
\psi_ N^2 
 + \int_0^1\frac{\dd x}{2x(1-x)}\int\frac{\dd^2 k_\perp}{(2\pi)^3} \frac{g^2}{(s_2-m^2)^2} 
 = 1.
\end{equation}
We can define
\begin{equation}
\begin{split}
I_\pi \equiv\,& 
\int_0^1\frac{\dd x}{2x(1-x)}\int\frac{\dd^2 k_\perp}{(2\pi)^3} \frac{g^2}{(s_2-m^2)^2}, \\
=\,& \frac{\alpha}{\pi} \int \dd x \, \frac{x(1-x)}{(1-x)\frac{\mu^2}{m^2}+x^2} 
\end{split}
\end{equation}

\begin{equation}
\psi_ N = \sqrt{1 - I_\pi}
\end{equation}

\begin{figure}
\centering
\includegraphics[width=0.6\textwidth]{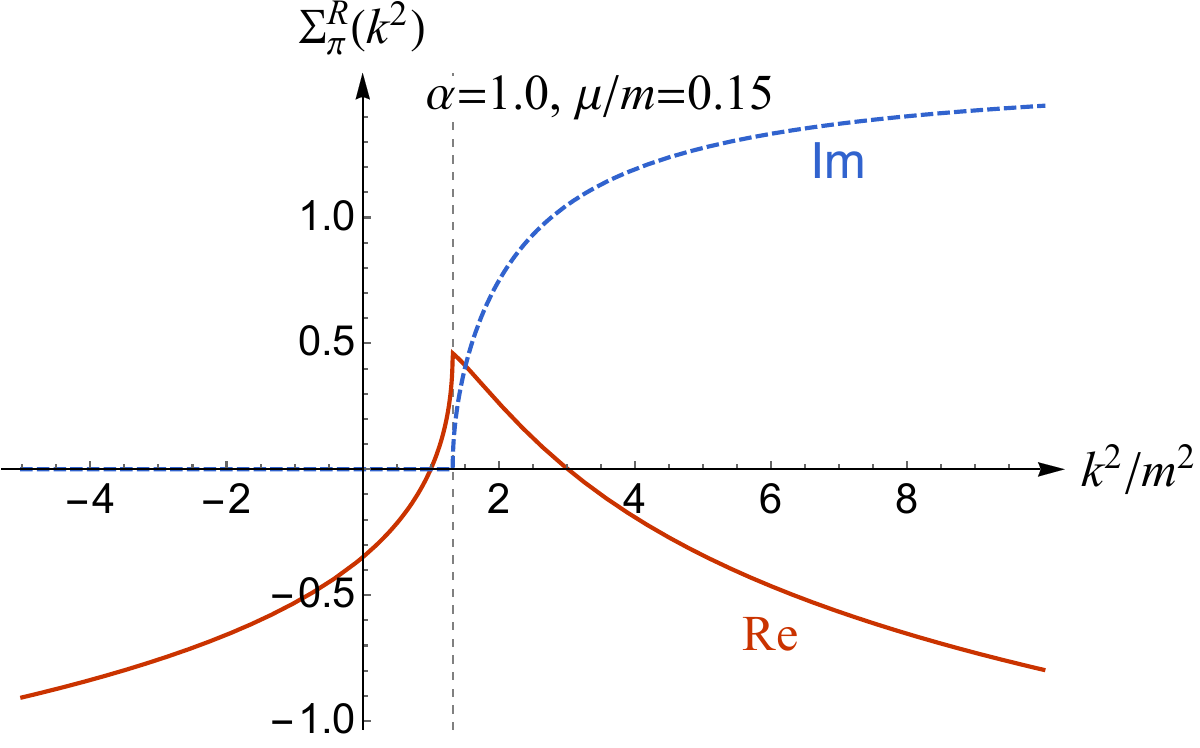}
\caption{The self-energy function $\Sigma_{\pi}^R(k^2)$ in two-body truncation.
The branch point is located at the pair creation 
threshold $k_\mathrm{th}^2 = (m+\mu)^2$, as indicated by the vertical dashed line.}
\label{fig:SE_Sigma}
\end{figure}

It is useful to introduce two auxiliary functions:
\begin{equation}
\begin{split}
f(\lambda) \equiv\,& \int_0^1\dd x \frac{x(1-x)}{(1-x)\lambda^2 + x^2} \\
=\,& (\lambda^2-1)\ln \lambda + \frac{\lambda(3-\lambda^2)}{\sqrt{4-\lambda^2}}\arctan\frac{\sqrt{4-\lambda^2}}{\lambda} - 1 \qquad (0 < \lambda < 2)
\end{split}
\end{equation}
and 
\begin{equation}
\begin{split}
F(\lambda, \rho)
\equiv\,& -\int_0^1 \dd x \ln\Big[(1-x)\lambda^2 + x - x(1-x)\rho\Big] \\
=\,& 
\begin{cases}
2 + \frac{1-\lambda^2-\rho-\sqrt{\Delta}}{2\rho} \ln \lambda^2 
+ \frac{\sqrt{\Delta}}{\rho} \ln \frac{1+\lambda^2-\rho+\sqrt{\Delta}}{2},
& \rho < (1-\lambda)^2, \\
1 + \frac{\lambda^2}{1-\lambda^2}\ln\lambda^2, & \rho=0
\end{cases}
\end{split}
\end{equation}
where $\Delta \equiv (1-\lambda^2-\rho)^2 - 4\lambda^2\rho$.
$\rho=0$ is a removable singularity.
They are related by: 
\begin{equation}
\frac{\partial }{\partial \rho}F(\lambda, \rho)\Big|_{\rho \to 1} = f(\lambda).
\end{equation}
With the help of these two auxiliary functions, we have,
\begin{align}
I_\pi = Z_\pi =\,& 1 - \frac{\alpha}{\pi}f\big(\mathsmaller{\frac{\mu}{m}}\big), \\
\Sigma_\pi^R(p^2) = \,& \frac{m^2}{Z_\pi} \frac{\alpha}{\pi} \Big[
F\big( \mathsmaller{\frac{\mu}{m}}, \mathsmaller{\frac{p^2}{m^2}} \big)
 - F\big( \mathsmaller{ \frac{\mu}{m} }, 1\big) \Big], \\
 g_\pi =\,& \frac{g}{\sqrt{Z_\pi}}
\end{align}
The self-energy $\Sigma_{ N\bar N}^R$ is shown in Fig.~\ref{fig:SE_Sigma}.

\subsection{Pion sector}

\begin{figure} \centering
\includegraphics[width=0.65\textwidth]{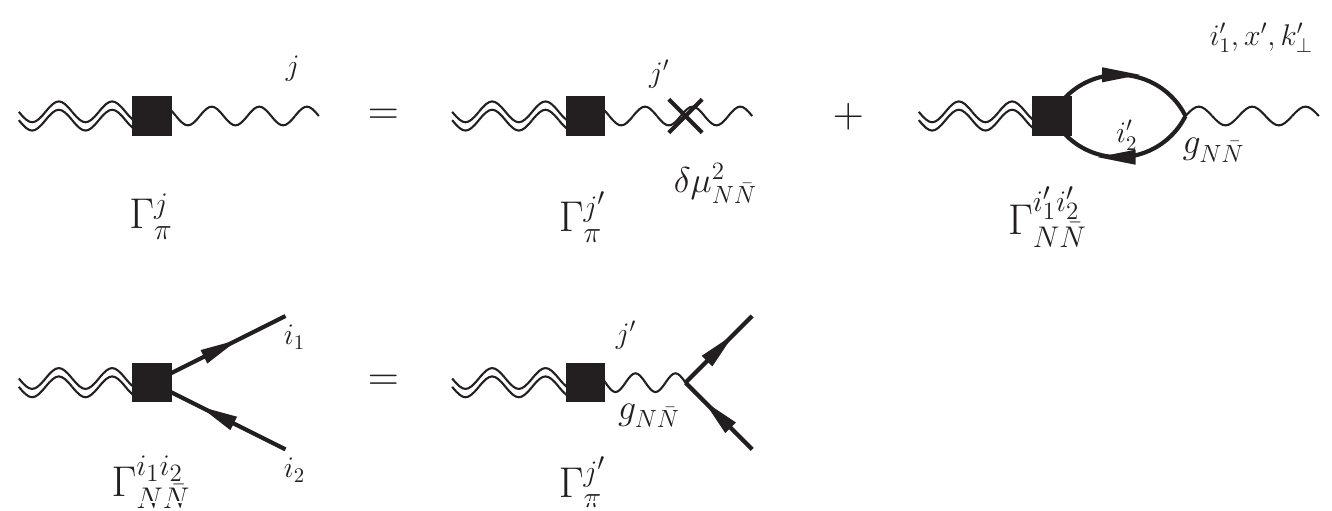}
\caption{Diagrammatic representation of the equation (\ref{eqn:Gamma_chi_barchi_N2}). The double wavy lines represent the physical pion. The solid straight lines represent the nucleon or anti-nucleon. The wavy lines represent the pion.}
\label{fig:gamma_chi_barchi_N2}
\end{figure}

The renormalization parameters were previously obtained by Smirnov \cite{Smirnov:2010}.
Here we present them for completeness. 
These parameters are determined from the physical pion sector with a truncation up to $|\pi\rangle + | N\bar N\rangle$, as shown in 
 Fig.~\ref{fig:gamma_chi_barchi_N2}.
 
\begin{align}\label{eqn:Gamma_chi_barchi_N2}
& \Gamma_\pi^j = \delta \mu^2_{ N\bar N} \sum_{j'}(-1)^{j'} \frac{\Gamma_\pi^{j'}}{\mu^2_{j'} - \mu^2}
+ \sum_{i_1', i'_2}(-1)^{i'_1+i'_2}\int_0^1\frac{\dd x'}{2x'(1-x')} \int\frac{\dd^2k'_\perp}{(2\pi)^3}
\frac{g_{ N\bar N} \Gamma_{ N\bar N}^{i_1'i'_2}(x', k'_\perp)}{s_2^{i'_1i'_2}-\mu^2}, \\
& \Gamma_{ N\bar N}^{i_1i_2} = g_{ N\bar N} \sum_{j'} (-1)^{j'} \frac{\Gamma_\pi^{j'}}{\mu^2_{j'} - \mu^2}.
\end{align}

The renormalization condition is, 
\begin{equation}
\Gamma^{i_1=0, i_2=0}_{ N\bar N}(x^\star, k^\star_\perp) = g
\quad \Big(\frac{k^{\star2}_\perp+m^2}{x^\star} + \frac{k^{\star2}_\perp+m^2}{1-x^\star} = \mu^2\Big)
\end{equation}

\begin{figure}
\centering
\includegraphics[width=0.6\textwidth]{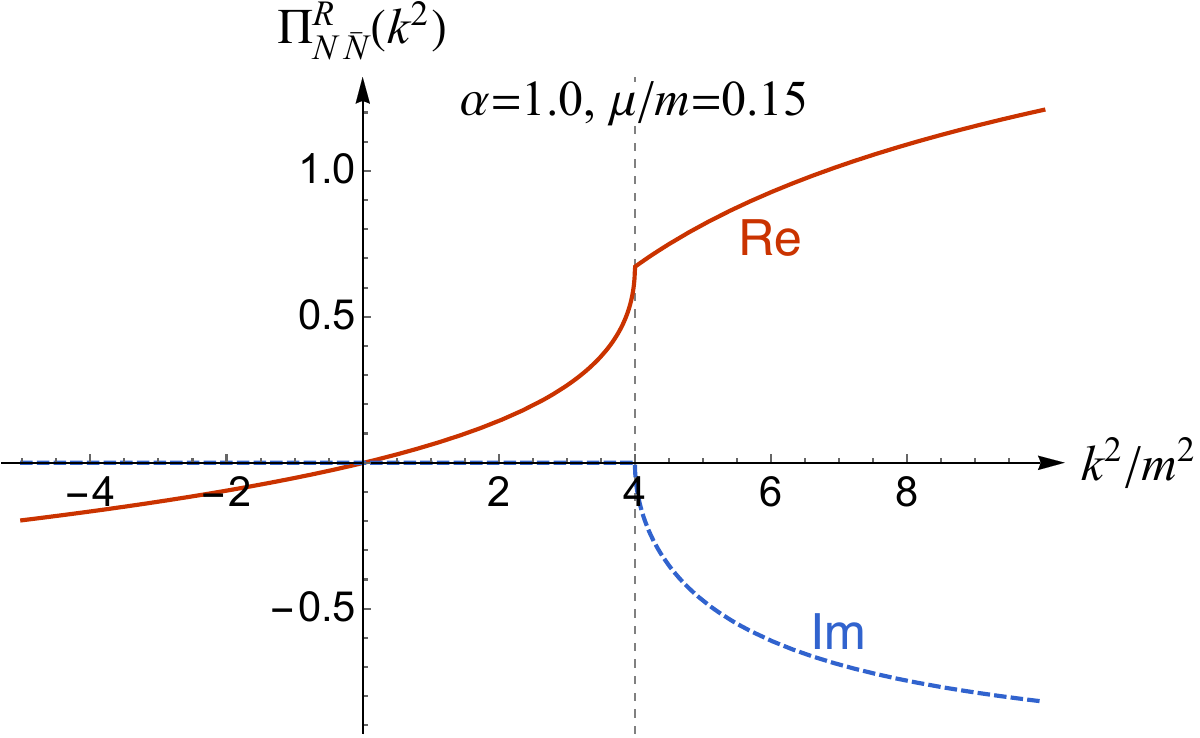}
\caption{The self-energy function $\Pi_{ N\bar N}^R(k^2)$  in two-body truncation.
The branch point is located at the pair creation 
threshold $k_\mathrm{th}^2 = 4m^2$, as indicated by the vertical dashed line.}
\label{fig:SE_Pi}
\end{figure}

Introduce two auxiliary functions,
\begin{equation}
\begin{split}
\phi(\lambda) =\,& \int_0^1 \dd x \frac{x(1-x)}{1 - x(1-x)\lambda^2} \\
=\,& \frac{4}{\lambda^3\sqrt{4-\lambda^2}}\arctan\frac{\lambda}{\sqrt{4-\lambda^2}} - \frac{1}{\lambda^2}
\end{split}
\end{equation} 
\begin{equation}
\begin{split}
\Phi(\rho) =\,& -\int_0^1 \dd x \ln \big[ 1 - x(1-x) \rho \big] \\
=\,& 
\begin{cases}
2 - 2 \sqrt{\frac{4-\rho}{\rho}} \mathrm{arctan}\sqrt{\frac{\rho}{4-\rho}}, & 0 \le \rho < 4; \\
2 - 2 \sqrt{\frac{4-\rho}{-\rho}} \mathrm{arctanh}\sqrt{\frac{-\rho}{4-\rho}}, & \rho < 0; \\
0, & \rho = 0.
\end{cases}
\end{split}
\end{equation}
$\rho = 0$ is a removable singularity. 
It is easy to show that they are related by the derivative, 
\begin{equation}
\Phi'(\lambda^2) = \phi(\lambda).
\end{equation}
With these two auxiliary functions, the renormalization parameters are, 
\begin{align}
Z_{ N\bar N} =\,& 1 - \frac{\alpha}{\pi}\phi(\mathsmaller{\frac{\mu}{m}}), \\
\Pi^R_{ N\bar N}(k^2) =\,& \frac{m^2}{Z_{ N\bar N}}\frac{\alpha}{\pi} \big[ \Phi(\mathsmaller{\frac{k^2}{m^2}}) - \Phi(\mathsmaller{\frac{\mu^2}{m^2}})\big], \\
g_{ N\bar N}=\,& \frac{g}{\sqrt{Z_{ N\bar N}}}.
\end{align}
The self-energy $\Pi_{ N\bar N}^R$ is shown in Fig.~\ref{fig:SE_Pi}.

\begin{thebibliography}{99}

\bibitem{Gross:2022hyw}
F.~Gross, E.~Klempt, S.~J.~Brodsky, A.~J.~Buras, V.~D.~Burkert, G.~Heinrich, K.~Jakobs, C.~A.~Meyer, K.~Orginos and M.~Strickland, \textit{et al.}
Eur. Phys. J. C \textbf{83}, 1125 (2023)
doi:10.1140/epjc/s10052-023-11949-2
[arXiv:2212.11107 [hep-ph]].

\bibitem{Gao:2021sml}
H.~Gao and M.~Vanderhaeghen,
Rev. Mod. Phys. \textbf{94}, no.1, 015002 (2022)
doi:10.1103/RevModPhys.94.015002
[arXiv:2105.00571 [hep-ph]].

\bibitem{Perdrisat:2006hj}
C.~F.~Perdrisat, V.~Punjabi and M.~Vanderhaeghen,
Prog. Part. Nucl. Phys. \textbf{59}, 694-764 (2007)
doi:10.1016/j.ppnp.2007.05.001
[arXiv:hep-ph/0612014 [hep-ph]].

\bibitem{Punjabi:2015bba}
V.~Punjabi, C.~F.~Perdrisat, M.~K.~Jones, E.~J.~Brash and C.~E.~Carlson,
Eur. Phys. J. A \textbf{51}, 79 (2015)
doi:10.1140/epja/i2015-15079-x
[arXiv:1503.01452 [nucl-ex]].

\bibitem{Pacetti:2014jai}
S.~Pacetti, R.~Baldini Ferroli and E.~Tomasi-Gustafsson,
Phys. Rept. \textbf{550-551}, 1-103 (2015)
doi:10.1016/j.physrep.2014.09.005

\bibitem{Miller:2010nz}
G.~A.~Miller,
Ann. Rev. Nucl. Part. Sci. \textbf{60}, 1-25 (2010)
doi:10.1146/annurev.nucl.012809.104508
[arXiv:1002.0355 [nucl-th]].

\bibitem{Hand:1963zz}
L.~N.~Hand, D.~G.~Miller and R.~Wilson,
Rev. Mod. Phys. \textbf{35}, 335 (1963)
doi:10.1103/RevModPhys.35.335

\bibitem{Hofstadter:1956qs}
R.~Hofstadter,
Rev. Mod. Phys. \textbf{28}, 214-254 (1956)
doi:10.1103/RevModPhys.28.214

\bibitem{Alexandrou:2012da}
C.~Alexandrou, C.~N.~Papanicolas and M.~Vanderhaeghen,
Rev. Mod. Phys. \textbf{84}, 1231 (2012)
doi:10.1103/RevModPhys.84.1231
[arXiv:1201.4511 [hep-ph]].

\bibitem{Lepage:1980fj}
G.~P.~Lepage and S.~J.~Brodsky,
Phys. Rev. D \textbf{22}, 2157 (1980)
doi:10.1103/PhysRevD.22.2157

\bibitem{Isgur:1984jm}
N.~Isgur and C.~H.~Llewellyn Smith,
Phys. Rev. Lett. \textbf{52}, 1080 (1984)
doi:10.1103/PhysRevLett.52.1080

\bibitem{Isgur:1988iw}
N.~Isgur and C.~H.~Llewellyn Smith,
Nucl. Phys. B \textbf{317}, 526-572 (1989)
doi:10.1016/0550-3213(89)90532-4

\bibitem{Isgur:1989cy}
N.~Isgur and C.~H.~Llewellyn Smith,
Phys. Lett. B \textbf{217}, 535-538 (1989)
doi:10.1016/0370-2693(89)90092-0

\bibitem{Sterman:1997sx}
G.~F.~Sterman and P.~Stoler,
Ann. Rev. Nucl. Part. Sci. \textbf{47}, 193-233 (1997)
doi:10.1146/annurev.nucl.47.1.193
[arXiv:hep-ph/9708370 [hep-ph]].

\bibitem{Bakker:2013cea}
B.~L.~G.~Bakker, A.~Bassetto, S.~J.~Brodsky, W.~Broniowski, S.~Dalley, T.~Frederico, S.~D.~Glazek, J.~R.~Hiller, C.~R.~Ji and V.~Karmanov, \textit{et al.}
Nucl. Phys. B Proc. Suppl. \textbf{251-252}, 165-174 (2014)
doi:10.1016/j.nuclphysbps.2014.05.004
[arXiv:1309.6333 [hep-ph]].

\bibitem{Fubini:1964boa}
S.~Fubini and G.~Furlan,
Physics Physique Fizika \textbf{1}, no.4, 229-247 (1965)
doi:10.1103/PhysicsPhysiqueFizika.1.229

\bibitem{Bjorken:1968dy}
J.~D.~Bjorken,
Phys. Rev. \textbf{179}, 1547-1553 (1969)
doi:10.1103/PhysRev.179.1547

\bibitem{Feynman:1969ej}
R.~P.~Feynman,
Phys. Rev. Lett. \textbf{23}, 1415-1417 (1969)
doi:10.1103/PhysRevLett.23.1415

\bibitem{Kogut:1972di}
J.~B.~Kogut and L.~Susskind,
Phys. Rept. \textbf{8}, 75-172 (1973)
doi:10.1016/0370-1573(73)90009-4

\bibitem{Altarelli:1977zs}
G.~Altarelli and G.~Parisi,
Nucl. Phys. B \textbf{126}, 298-318 (1977)
doi:10.1016/0550-3213(77)90384-4

\bibitem{Frankfurt:1977vc}
L.~L.~Frankfurt and M.~I.~Strikman,
Nucl. Phys. B \textbf{148}, 107-140 (1979)
doi:10.1016/0550-3213(79)90018-X

\bibitem{Altarelli:1981ax}
G.~Altarelli,
Phys. Rept. \textbf{81}, 1 (1982)
doi:10.1016/0370-1573(82)90127-2

\bibitem{Frankfurt:1981mk}
L.~L.~Frankfurt and M.~I.~Strikman,
Phys. Rept. \textbf{76}, 215-347 (1981)
doi:10.1016/0370-1573(81)90129-0

\bibitem{Chernyak:1983ej}
V.~L.~Chernyak and A.~R.~Zhitnitsky,
Phys. Rept. \textbf{112}, 173 (1984)
doi:10.1016/0370-1573(84)90126-1

\bibitem{Frankfurt:1988nt}
L.~L.~Frankfurt and M.~I.~Strikman,
Phys. Rept. \textbf{160}, 235-427 (1988)
doi:10.1016/0370-1573(88)90179-2

\bibitem{Frankfurt:1991rk}
L.~Frankfurt and M.~Strikman,
Prog. Part. Nucl. Phys. \textbf{27}, 135-193 (1991)
doi:10.1016/0146-6410(91)90004-8

\bibitem{Goeke:2001tz}
K.~Goeke, M.~V.~Polyakov and M.~Vanderhaeghen,
Prog. Part. Nucl. Phys. \textbf{47}, 401-515 (2001)
doi:10.1016/S0146-6410(01)00158-2
[arXiv:hep-ph/0106012 [hep-ph]].

\bibitem{Braun:2003rp}
V.~M.~Braun, G.~P.~Korchemsky and D.~M\"uller,
Prog. Part. Nucl. Phys. \textbf{51}, 311-398 (2003)
doi:10.1016/S0146-6410(03)90004-4
[arXiv:hep-ph/0306057 [hep-ph]].

\bibitem{Ivanov:2004ax}
I.~P.~Ivanov, N.~N.~Nikolaev and A.~A.~Savin,
Phys. Part. Nucl. \textbf{37}, 1-85 (2006)
doi:10.1134/S1063779606010011
[arXiv:hep-ph/0501034 [hep-ph]].

\bibitem{Kovchegov:2012mbw}
Y.~V.~Kovchegov and E.~Levin,
Camb. Monogr. Part. Phys. Nucl. Phys. Cosmol. \textbf{33}, 1-350 (2012)
Oxford University Press, 2013,
ISBN 978-1-009-29144-6, 978-1-009-29141-5, 978-1-009-29142-2, 978-0-521-11257-4, 978-1-139-55768-9
doi:10.1017/9781009291446

\bibitem{Cruz-Santiago:2015dla}
C.~Cruz-Santiago, P.~Kotko and A.~M.~Sta\'sto,
Prog. Part. Nucl. Phys. \textbf{85}, 82-131 (2015)
doi:10.1016/j.ppnp.2015.07.002

\bibitem{Sterman:2016etx}
G.~Sterman,
Int. J. Mod. Phys. A \textbf{31}, no.09, 1630005 (2016)
doi:10.1142/S0217751X16300052
[arXiv:1602.02307 [hep-ph]].

\bibitem{Dirac:1949cp}
P.~A.~M.~Dirac,
Rev. Mod. Phys. \textbf{21}, 392-399 (1949)
doi:10.1103/RevModPhys.21.392

\bibitem{Weinberg:1966jm}
S.~Weinberg,
Phys. Rev. \textbf{150}, 1313-1318 (1966)
doi:10.1103/PhysRev.150.1313

\bibitem{Susskind:1967rg}
L.~Susskind,
Phys. Rev. \textbf{165}, 1535-1546 (1968)
doi:10.1103/PhysRev.165.1535

\bibitem{Kogut:1969xa}
J.~B.~Kogut and D.~E.~Soper,
Phys. Rev. D \textbf{1}, 2901-2913 (1970)
doi:10.1103/PhysRevD.1.2901

\bibitem{Bjorken:1970ah}
J.~D.~Bjorken, J.~B.~Kogut and D.~E.~Soper,
Phys. Rev. D \textbf{3}, 1382 (1971)
doi:10.1103/PhysRevD.3.1382

\bibitem{Brodsky:1973kb}
S.~J.~Brodsky, R.~Roskies and R.~Suaya,
Phys. Rev. D \textbf{8}, 4574 (1973)
doi:10.1103/PhysRevD.8.4574

\bibitem{Coester:1992cg}
F.~Coester,
Prog. Part. Nucl. Phys. \textbf{29}, 1-32 (1992)
doi:10.1016/0146-6410(92)90002-J

\bibitem{Fuda:1992uh}
M.~G.~Fuda,
Nucl. Phys. A \textbf{543}, 111C-126C (1992)
doi:10.1016/0375-9474(92)90414-F

\bibitem{Zhang:1994ti}
W.~M.~Zhang,
Chin. J. Phys. \textbf{32}, 717-808 (1994)
[arXiv:hep-ph/9412244 [hep-ph]].

\bibitem{Burkardt:1995ct}
M.~Burkardt,
Adv. Nucl. Phys. \textbf{23}, 1-74 (1996)
doi:10.1007/0-306-47067-5\_1
[arXiv:hep-ph/9505259 [hep-ph]].

\bibitem{Brodsky:1997de}
S.~J.~Brodsky, H.~C.~Pauli and S.~S.~Pinsky,
Phys. Rept. \textbf{301}, 299-486 (1998)
doi:10.1016/S0370-1573(97)00089-6
[arXiv:hep-ph/9705477 [hep-ph]].

\bibitem{Carbonell:1998rj}
J.~Carbonell, B.~Desplanques, V.~A.~Karmanov and J.~F.~Mathiot,
Phys. Rept. \textbf{300}, 215-347 (1998)
doi:10.1016/S0370-1573(97)00090-2
[arXiv:nucl-th/9804029 [nucl-th]].

\bibitem{Miller:2000kv}
G.~A.~Miller,
Prog. Part. Nucl. Phys. \textbf{45}, 83-155 (2000)
doi:10.1016/S0146-6410(00)00103-4
[arXiv:nucl-th/0002059 [nucl-th]].

\bibitem{Heinzl:2000ht}
T.~Heinzl,
Lect. Notes Phys. \textbf{572}, 55-142 (2001)
doi:10.1007/3-540-45114-5\_2
[arXiv:hep-th/0008096 [hep-th]].

\bibitem{Garsevanishvili:2007jp}
V.~R.~Garsevanishvili, A.~A.~Khelashvili, Z.~R.~Menteshashvili and M.~S.~Nioradze,
Phys. Rept. \textbf{458}, 247-300 (2008)
doi:10.1016/j.physrep.2007.11.002
[arXiv:0705.3630 [hep-th]].

\bibitem{Martinovic:2007nee}
\v{L}.~Martinovi\v{c},
Acta Phys. Slov. \textbf{57}, no.3, 407-564 (2007)

\bibitem{Frederico:2010zh}
T.~Frederico and G.~Salme,
Few Body Syst. \textbf{49}, 163-175 (2011)
doi:10.1007/s00601-010-0163-z
[arXiv:1011.1850 [nucl-th]].

\bibitem{Brodsky:2014yha}
S.~J.~Brodsky, G.~F.~de Teramond, H.~G.~Dosch and J.~Erlich,
Phys. Rept. \textbf{584}, 1-105 (2015)
doi:10.1016/j.physrep.2015.05.001
[arXiv:1407.8131 [hep-ph]].

\bibitem{Bakker:2014cua}
B.~L.~G.~Bakker and C.~R.~Ji,
Prog. Part. Nucl. Phys. \textbf{74}, 1-34 (2014)
doi:10.1016/j.ppnp.2013.10.001

\bibitem{Hiller:2016itl}
J.~R.~Hiller,
Prog. Part. Nucl. Phys. \textbf{90}, 75-124 (2016)
doi:10.1016/j.ppnp.2016.06.002
[arXiv:1606.08348 [hep-ph]].

\bibitem{Ji:2020ect}
X.~Ji, Y.~S.~Liu, Y.~Liu, J.~H.~Zhang and Y.~Zhao,
Rev. Mod. Phys. \textbf{93}, no.3, 035005 (2021)
doi:10.1103/RevModPhys.93.035005
[arXiv:2004.03543 [hep-ph]].

\bibitem{Drell:1969km}
S.~D.~Drell and T.~M.~Yan,
Phys. Rev. Lett. \textbf{24}, 181-185 (1970)
doi:10.1103/PhysRevLett.24.181

\bibitem{West:1970av}
G.~B.~West,
Phys. Rev. Lett. \textbf{24}, 1206-1209 (1970)
doi:10.1103/PhysRevLett.24.1206

\bibitem{Sawicki:1992qj}
M.~Sawicki,
Phys. Rev. D \textbf{46}, 474-477 (1992)
doi:10.1103/PhysRevD.46.474

\bibitem{Demchuk:1995zx}
N.~B.~Demchuk, I.~L.~Grach, I.~M.~Narodetski and S.~Simula,
Phys. Atom. Nucl. \textbf{59}, 2152-2163 (1996)
[arXiv:hep-ph/9601369 [hep-ph]].

\bibitem{deMelo:1997hh}
J.~P.~B.~C.~de Melo and T.~Frederico,
Phys. Rev. C \textbf{55}, 2043 (1997)
doi:10.1103/PhysRevC.55.2043
[arXiv:nucl-th/9706032 [nucl-th]].

\bibitem{deMelo:1997cb}
J.~P.~C.~B.~de Melo, H.~W.~L.~Naus and T.~Frederico,
Phys. Rev. C \textbf{59}, 2278-2281 (1999)
doi:10.1103/PhysRevC.59.2278
[arXiv:hep-ph/9710228 [hep-ph]].

\bibitem{deMelo:1998an}
J.~P.~B.~C.~de Melo, J.~H.~O.~Sales, T.~Frederico and P.~U.~Sauer,
Nucl. Phys. A \textbf{631}, 574C-579C (1998)
doi:10.1016/S0375-9474(98)00070-0
[arXiv:hep-ph/9802325 [hep-ph]].

\bibitem{Choi:1998nf}
H.~M.~Choi and C.~R.~Ji,
Phys. Rev. D \textbf{58}, 071901 (1998)
doi:10.1103/PhysRevD.58.071901
[arXiv:hep-ph/9805438 [hep-ph]].

\bibitem{deMelo:1999gn}
J.~P.~B.~C.~de Melo, T.~Frederico, H.~W.~L.~Naus and P.~U.~Sauer,
Nucl. Phys. A \textbf{660}, 219-231 (1999)
doi:10.1016/S0375-9474(99)00371-1
[arXiv:hep-ph/9908384 [hep-ph]].

\bibitem{Bakker:2000pk}
B.~L.~G.~Bakker, H.~M.~Choi and C.~R.~Ji,
Phys. Rev. D \textbf{63}, 074014 (2001)
doi:10.1103/PhysRevD.63.074014
[arXiv:hep-ph/0008147 [hep-ph]].

\bibitem{Tiburzi:2001je}
B.~C.~Tiburzi and G.~A.~Miller,
Phys. Rev. D \textbf{65}, 074009 (2002)
doi:10.1103/PhysRevD.65.074009
[arXiv:hep-ph/0109174 [hep-ph]].

\bibitem{Melikhov:2001pm}
D.~Melikhov and S.~Simula,
Phys. Rev. D \textbf{65}, 094043 (2002)
doi:10.1103/PhysRevD.65.094043
[arXiv:hep-ph/0112044 [hep-ph]].

\bibitem{Simula:2002vm}
S.~Simula,
Phys. Rev. C \textbf{66}, 035201 (2002)
doi:10.1103/PhysRevC.66.035201
[arXiv:nucl-th/0204015 [nucl-th]].

\bibitem{deMelo:2002yq}
J.~P.~B.~C.~de Melo, T.~Frederico, E.~Pace and G.~Salme,
Nucl. Phys. A \textbf{707}, 399-424 (2002)
doi:10.1016/S0375-9474(02)00990-9
[arXiv:nucl-th/0205010 [nucl-th]].

\bibitem{Tiburzi:2002mn}
B.~C.~Tiburzi and G.~A.~Miller,
[arXiv:hep-ph/0205109 [hep-ph]].

\bibitem{Melikhov:2002mp}
D.~Melikhov and S.~Simula,
Phys. Lett. B \textbf{556}, 135-141 (2003)
doi:10.1016/S0370-2693(03)00124-2
[arXiv:hep-ph/0211277 [hep-ph]].

\bibitem{deMelo:2003uk}
J.~P.~B.~C.~de Melo, T.~Frederico, E.~Pace and G.~Salme,
Phys. Lett. B \textbf{581}, 75-81 (2004)
doi:10.1016/j.physletb.2003.11.072
[arXiv:hep-ph/0311369 [hep-ph]].

\bibitem{Tiburzi:2004ye}
B.~C.~Tiburzi,
[arXiv:nucl-th/0407005 [nucl-th]].

\bibitem{deMelo:2005cy}
J.~P.~B.~C.~de Melo, T.~Frederico, E.~Pace and G.~Salme,
Phys. Rev. D \textbf{73}, 074013 (2006)
doi:10.1103/PhysRevD.73.074013
[arXiv:hep-ph/0508001 [hep-ph]].

\bibitem{Suzuki:2012ni}
A.~T.~Suzuki, J.~H.~Sales and L.~A.~Soriano,
Phys. Rev. D \textbf{88}, no.2, 025036 (2013)
doi:10.1103/PhysRevD.88.025036
[arXiv:1209.4681 [hep-th]].

\bibitem{deMelo:2012hj}
J.~P.~B.~C.~de Melo and T.~Frederico,
Phys. Lett. B \textbf{708}, 87-92 (2012)
doi:10.1016/j.physletb.2012.01.021
[arXiv:1202.0734 [hep-ph]].

\bibitem{Eichmann:2016yit}
G.~Eichmann, H.~Sanchis-Alepuz, R.~Williams, R.~Alkofer and C.~S.~Fischer,
Prog. Part. Nucl. Phys. \textbf{91}, 1-100 (2016)
doi:10.1016/j.ppnp.2016.07.001
[arXiv:1606.09602 [hep-ph]].

\bibitem{Maris:2003vk}
P.~Maris and C.~D.~Roberts,
Int. J. Mod. Phys. E \textbf{12}, 297-365 (2003)
doi:10.1142/S0218301303001326
[arXiv:nucl-th/0301049 [nucl-th]].

\bibitem{Carbonell:2017isq}
J.~Carbonell, T.~Frederico and V.~A.~Karmanov,
Eur. Phys. J. C \textbf{77}, no.1, 58 (2017)
doi:10.1140/epjc/s10052-017-4616-0
[arXiv:1701.02479 [hep-ph]].

\bibitem{Bakker:2002mt}
B.~L.~G.~Bakker, H.~M.~Choi and C.~R.~Ji,
Phys. Rev. D \textbf{65}, 116001 (2002)
doi:10.1103/PhysRevD.65.116001
[arXiv:hep-ph/0202217 [hep-ph]].

\bibitem{Choi:2004ww}
H.~M.~Choi and C.~R.~Ji,
Phys. Rev. D \textbf{70}, 053015 (2004)
doi:10.1103/PhysRevD.70.053015
[arXiv:hep-ph/0402114 [hep-ph]].

\bibitem{Choi:2005fj}
H.~M.~Choi and C.~R.~Ji,
Phys. Rev. D \textbf{72}, 013004 (2005)
doi:10.1103/PhysRevD.72.013004
[arXiv:hep-ph/0504219 [hep-ph]].

\bibitem{He:2005hw}
J.~He and Y.~b.~Dong,
J. Phys. G \textbf{32}, 189-202 (2006)
doi:10.1088/0954-3899/32/2/010
[arXiv:hep-ph/0512056 [hep-ph]].

\bibitem{Arifi:2022qnd}
A.~J.~Arifi, H.~M.~Choi, C.~R.~Ji and Y.~Oh,
Phys. Rev. D \textbf{107}, no.5, 053003 (2023)
doi:10.1103/PhysRevD.107.053003
[arXiv:2210.12780 [hep-ph]].

\bibitem{Brodsky:1998hn}
S.~J.~Brodsky and D.~S.~Hwang,
Nucl. Phys. B \textbf{543}, 239-252 (1999)
doi:10.1016/S0550-3213(98)00807-4
[arXiv:hep-ph/9806358 [hep-ph]].

\bibitem{Li:2017uug}
Y.~Li, P.~Maris and J.~Vary,
Phys. Rev. D \textbf{97}, no.5, 054034 (2018)
doi:10.1103/PhysRevD.97.054034
[arXiv:1712.03467 [hep-ph]].

\bibitem{Li:2019kpr}
M.~Li, Y.~Li, P.~Maris and J.~P.~Vary,
Phys. Rev. D \textbf{100}, no.3, 036006 (2019)
doi:10.1103/PhysRevD.100.036006
[arXiv:1906.07306 [nucl-th]].

\bibitem{Brodsky:2007hb}
S.~J.~Brodsky and G.~F.~de Teramond,
Phys. Rev. D \textbf{77}, 056007 (2008)
doi:10.1103/PhysRevD.77.056007
[arXiv:0707.3859 [hep-ph]].

\bibitem{Brodsky:2008pf}
S.~J.~Brodsky and G.~F.~de Teramond,
Phys. Rev. D \textbf{78}, 025032 (2008)
doi:10.1103/PhysRevD.78.025032
[arXiv:0804.0452 [hep-ph]].

\bibitem{Li:2023izn}
Y.~Li and J.~P.~Vary,
Phys. Rev. D \textbf{109}, no.5, L051501 (2024)
doi:10.1103/PhysRevD.109.L051501
[arXiv:2312.02543 [hep-th]].

\bibitem{Lev:1994au}
F.~M.~Lev,
Annals Phys. \textbf{237}, 355-419 (1995)
doi:10.1006/aphy.1995.1013
[arXiv:hep-ph/9403222 [hep-ph]].

\bibitem{Lev:1998qz}
F.~M.~Lev, E.~Pace and G.~Salme,
Nucl. Phys. A \textbf{641}, 229-259 (1998)
doi:10.1016/S0375-9474(98)00469-2
[arXiv:hep-ph/9807255 [hep-ph]].

\bibitem{Lev:1999me}
F.~M.~Lev, E.~Pace and G.~Salme,
Phys. Rev. Lett. \textbf{83}, 5250-5253 (1999)
doi:10.1103/PhysRevLett.83.5250
[arXiv:nucl-th/9910049 [nucl-th]].

\bibitem{Lev:2000vm}
F.~M.~Lev, E.~Pace and G.~Salme,
Phys. Rev. C \textbf{62}, 064004 (2000)
doi:10.1103/PhysRevC.62.064004
[arXiv:nucl-th/0006053 [nucl-th]].

\bibitem{Lev:1999au}
F.~M.~Lev, E.~Pace and G.~Salme,
Nucl. Phys. A \textbf{663}, 365-368 (2000)
doi:10.1016/S0375-9474(99)00618-1
[arXiv:nucl-th/9909027 [nucl-th]].

\bibitem{Pace:2001tzb}
E.~Pace and G.~Salm\`e,
Nucl. Phys. A \textbf{684}, 487-489 (2001)
doi:10.1016/S0375-9474(01)00472-9

\bibitem{Pace:2006yj}
E.~Pace, G.~Salme, T.~Frederico, S.~Pisano and J.~P.~B.~C.~de Melo,
Nucl. Phys. A \textbf{790}, 606-609 (2007)
doi:10.1016/j.nuclphysa.2007.03.130
[arXiv:hep-ph/0611328 [hep-ph]].

\bibitem{deMelo:2006bs}
J.~P.~B.~C.~de Melo, T.~Frederico, E.~Pace, S.~Pisano and G.~Salme,
Nucl. Phys. A \textbf{782}, 69-76 (2007)
doi:10.1016/j.nuclphysa.2006.10.027
[arXiv:hep-ph/0607342 [hep-ph]].

\bibitem{Marinho:2007zzb}
J.~A.~O.~Marinho, T.~Frederico and P.~U.~Sauer,
Phys. Rev. D \textbf{76}, 096001 (2007)
doi:10.1103/PhysRevD.76.096001

\bibitem{Pace:2008obp}
E.~Pace, G.~Salme, T.~Frederico and S.~Pisano,
Few-Body Systems \textbf{44}, no.1-4, 299-302 (2008)
doi:10.1007/s00601-008-0313-8

\bibitem{Huang:2008jd}
Y.~Huang and W.~N.~Polyzou,
Phys. Rev. C \textbf{80}, 025503 (2009)
doi:10.1103/PhysRevC.80.025503
[arXiv:0812.2180 [nucl-th]].

\bibitem{Pace:2011zz}
E.~Pace, J.~A.~Marinho, G.~Salme and T.~Frederico,
Few Body Syst. \textbf{50}, 431-434 (2011)
doi:10.1007/s00601-010-0159-8

\bibitem{Polyzou:2023ldi}
W.~Polyzou,
Few Body Syst. \textbf{65}, no.1, 2 (2024)
doi:10.1007/s00601-023-01871-4
[arXiv:2310.19243 [nucl-th]].

\bibitem{Chabysheva:2009vm}
S.~S.~Chabysheva and J.~R.~Hiller,
Phys. Rev. D \textbf{81}, 074030 (2010)
doi:10.1103/PhysRevD.81.074030
[arXiv:0911.4455 [hep-ph]].

\bibitem{Li:2014kfa}
Y.~Li, V.~A.~Karmanov, P.~Maris and J.~P.~Vary,
Few Body Syst. \textbf{56}, no.6-9, 495-501 (2015)
doi:10.1007/s00601-015-0965-0
[arXiv:1411.1707 [nucl-th]].

\bibitem{Li:2015iaw}
Y.~Li, V.~A.~Karmanov, P.~Maris and J.~P.~Vary,
Phys. Lett. B \textbf{748}, 278-283 (2015)
doi:10.1016/j.physletb.2015.07.014
[arXiv:1504.05233 [nucl-th]].

\bibitem{Xu:2024sjt}
S.~Xu, Y.~Liu, C.~Mondal, J.~Lan, X.~Zhao, Y.~Li and J.~P.~Vary,
[arXiv:2408.11298 [hep-ph]].

\bibitem{Cao:2024fto}
X.~Cao, Y.~Li and J.~P.~Vary,
Phys. Rev. D \textbf{110}, no.7, 076025 (2024)
doi:10.1103/PhysRevD.110.076025
[arXiv:2408.09535 [hep-ph]].

\bibitem{Duan:2024dhy}
Y.~Duan, S.~Xu, S.~Cheng, X.~Zhao, Y.~Li and J.~P.~Vary,
[arXiv:2404.07755 [hep-ph]].

\bibitem{Cao:2024rul}
X.~Cao, S.~Xu, Y.~Li, G.~Chen, X.~Zhao, V.~A.~Karmanov and J.~P.~Vary,
JHEP \textbf{07}, 095 (2024)
doi:10.1007/JHEP07(2024)095
[arXiv:2405.06896 [hep-ph]].

\bibitem{Pauli:1985pv}
H.~C.~Pauli and S.~J.~Brodsky,
Phys. Rev. D \textbf{32}, 1993 (1985)
doi:10.1103/PhysRevD.32.1993

\bibitem{Pauli:1985ps}
H.~C.~Pauli and S.~J.~Brodsky,
Phys. Rev. D \textbf{32}, 2001 (1985)
doi:10.1103/PhysRevD.32.2001

\bibitem{Vary:2009gt}
J.~P.~Vary, H.~Honkanen, J.~Li, P.~Maris, S.~J.~Brodsky, A.~Harindranath, G.~F.~de Teramond, P.~Sternberg, E.~G.~Ng and C.~Yang,
Phys. Rev. C \textbf{81}, 035205 (2010)
doi:10.1103/PhysRevC.81.035205
[arXiv:0905.1411 [nucl-th]].

\bibitem{Brodsky:2005yu}
S.~J.~Brodsky, J.~R.~Hiller and G.~McCartor,
Annals Phys. \textbf{321}, 1240-1264 (2006)
doi:10.1016/j.aop.2005.09.005
[arXiv:hep-ph/0508295 [hep-ph]].

\bibitem{Baye:2015xoi}
D.~Baye,
Phys. Rept. \textbf{565}, 1-107 (2015)
doi:10.1016/j.physrep.2014.11.006

\bibitem{Baye:2002tix}
D.~Baye, M.~Hesse and M.~Vincke,
Phys. Rev. E \textbf{65}, no.2, 026701 (2002)
doi:10.1103/PhysRevE.65.026701

\bibitem{Leutwyler:1977vy}
H.~Leutwyler and J.~Stern,
Annals Phys. \textbf{112}, 94 (1978)
doi:10.1016/0003-4916(78)90082-9

\bibitem{Miller:2007uy}
G.~A.~Miller,
Phys. Rev. Lett. \textbf{99}, 112001 (2007)
doi:10.1103/PhysRevLett.99.112001
[arXiv:0705.2409 [nucl-th]].

\bibitem{Miller:2009sg}
G.~A.~Miller,
Phys. Rev. C \textbf{80}, 045210 (2009)
doi:10.1103/PhysRevC.80.045210
[arXiv:0908.1535 [nucl-th]].

\bibitem{Miller:2009qu}
G.~A.~Miller,
Phys. Rev. C \textbf{79}, 055204 (2009)
doi:10.1103/PhysRevC.79.055204
[arXiv:0901.1117 [nucl-th]].

\bibitem{Miller:2018ybm}
G.~A.~Miller,
Phys. Rev. C \textbf{99}, no.3, 035202 (2019)
doi:10.1103/PhysRevC.99.035202
[arXiv:1812.02714 [nucl-th]].

\bibitem{Freese:2021czn}
A.~Freese and G.~A.~Miller,
Phys. Rev. D \textbf{103}, 094023 (2021)
doi:10.1103/PhysRevD.103.094023
[arXiv:2102.01683 [hep-ph]].

\bibitem{Freese:2021qtb}
A.~Freese and G.~A.~Miller,
Phys. Rev. D \textbf{104}, no.1, 014024 (2021)
doi:10.1103/PhysRevD.104.014024
[arXiv:2104.03213 [hep-ph]].

\bibitem{Freese:2021mzg}
A.~Freese and G.~A.~Miller,
Phys. Rev. D \textbf{105}, no.1, 014003 (2022)
doi:10.1103/PhysRevD.105.014003
[arXiv:2108.03301 [hep-ph]].

\bibitem{Freese:2022fat}
A.~Freese and G.~A.~Miller,
Phys. Rev. D \textbf{108}, no.3, 034008 (2023)
doi:10.1103/PhysRevD.108.034008
[arXiv:2210.03807 [hep-ph]].

\bibitem{Li:2022hyf}
Y.~Li, W.~b.~Dong, Y.~l.~Yin, Q.~Wang and J.~P.~Vary,
Phys. Lett. B \textbf{838}, 137676 (2023)
doi:10.1016/j.physletb.2023.137676
[arXiv:2206.12903 [hep-ph]].

\bibitem{Freese:2023jcp}
A.~Freese and G.~A.~Miller,
Phys. Rev. D \textbf{107}, no.7, 074036 (2023)
doi:10.1103/PhysRevD.107.074036
[arXiv:2302.09171 [hep-ph]].

\bibitem{Freese:2023abr}
A.~Freese and G.~A.~Miller,
Phys. Rev. D \textbf{108}, no.9, 094026 (2023)
doi:10.1103/PhysRevD.108.094026
[arXiv:2307.11165 [hep-ph]].

\bibitem{Brodsky:2009zd}
S.~J.~Brodsky and R.~Shrock,
Proc. Nat. Acad. Sci. \textbf{108}, 45-50 (2011)
doi:10.1073/pnas.1010113107
[arXiv:0905.1151 [hep-th]].

\bibitem{Hornbostel:1991qj}
K.~Hornbostel,
Phys. Rev. D \textbf{45}, 3781-3801 (1992)
doi:10.1103/PhysRevD.45.3781

\bibitem{Heinzl:1991vd}
T.~Heinzl, S.~Krusche and E.~Werner,
Phys. Lett. B \textbf{256}, 55-59 (1991)
doi:10.1016/0370-2693(91)90218-F

\bibitem{Chakrabarti:2003tc}
D.~Chakrabarti, A.~Harindranath, L.~Martinovic, G.~B.~Pivovarov and J.~P.~Vary,
Phys. Lett. B \textbf{617}, 92-98 (2005)
doi:10.1016/j.physletb.2005.05.012
[arXiv:hep-th/0310290 [hep-th]].

\bibitem{Martinovic:2002bv}
L.~Martinovic,
Phys. Rev. D \textbf{78}, 105009 (2008)
doi:10.1103/PhysRevD.78.105009
[arXiv:hep-th/0207137 [hep-th]].

\bibitem{Chang:1968bh}
S.~J.~Chang and S.~K.~Ma,
Phys. Rev. \textbf{180}, 1506-1513 (1969)
doi:10.1103/PhysRev.180.1506

\bibitem{Yan:1973qg}
T.~M.~Yan,
Phys. Rev. D \textbf{7}, 1780-1800 (1973)
doi:10.1103/PhysRevD.7.1780

\bibitem{Chabysheva:2022duu}
S.~S.~Chabysheva and J.~R.~Hiller,
Phys. Rev. D \textbf{105}, no.11, 116006 (2022)
doi:10.1103/PhysRevD.105.116006
[arXiv:2201.00123 [hep-th]].

\bibitem{Ilderton:2014mla}
A.~Ilderton,
JHEP \textbf{09}, 166 (2014)
doi:10.1007/JHEP09(2014)166
[arXiv:1406.1513 [hep-th]].

\bibitem{Tomaras:2001vs}
T.~N.~Tomaras, N.~C.~Tsamis and R.~P.~Woodard,
JHEP \textbf{11}, 008 (2001)
doi:10.1088/1126-6708/2001/11/008
[arXiv:hep-th/0108090 [hep-th]].

\bibitem{Heinzl:2002uy}
T.~Heinzl,
[arXiv:hep-th/0212202 [hep-th]].

\bibitem{Burkardt:2000jn}
M.~Burkardt,
Nucl. Phys. A \textbf{670}, 72-75 (2000)
doi:10.1016/S0375-9474(00)00072-5

\bibitem{McCartor:1991ch}
G.~McCartor and D.~G.~Robertson,
Z. Phys. C \textbf{53}, 679-686 (1992)
doi:10.1007/BF01559747

\bibitem{Robertson:1992nj}
D.~G.~Robertson,
Phys. Rev. D \textbf{47}, 2549-2553 (1993)
doi:10.1103/PhysRevD.47.2549

\bibitem{Pinsky:1993yi}
S.~S.~Pinsky and B.~van de Sande,
Phys. Rev. D \textbf{49}, 2001-2013 (1994)
doi:10.1103/PhysRevD.49.2001
[arXiv:hep-ph/9310330 [hep-ph]].

\bibitem{Heinzl:1995xj}
T.~Heinzl, C.~Stern, E.~Werner and B.~Zellermann,
Z. Phys. C \textbf{72}, 353-364 (1996)
doi:10.1007/s002880050256
[arXiv:hep-th/9512179 [hep-th]].

\bibitem{Perry:1990mz}
R.~J.~Perry, A.~Harindranath and K.~G.~Wilson,
Phys. Rev. Lett. \textbf{65}, 2959-2962 (1990)
doi:10.1103/PhysRevLett.65.2959

\bibitem{Burkardt:1992sz}
M.~Burkardt,
Phys. Rev. D \textbf{47}, 4628-4633 (1993)
doi:10.1103/PhysRevD.47.4628

\bibitem{Fitzpatrick:2018ttk}
A.~L.~Fitzpatrick, J.~Kaplan, E.~Katz, L.~G.~Vitale and M.~T.~Walters,
JHEP \textbf{08}, 120 (2018)
doi:10.1007/JHEP08(2018)120
[arXiv:1803.10793 [hep-th]].

\bibitem{Collins:2018aqt}
J.~Collins,
[arXiv:1801.03960 [hep-ph]].

\bibitem{Martinovic:2018apr}
L.~Martinovic and A.~Dorokhov,
Phys. Lett. B \textbf{811}, 135925 (2020)
doi:10.1016/j.physletb.2020.135925
[arXiv:1812.02336 [hep-th]].

\bibitem{Brodsky:1998hs}
S.~J.~Brodsky, J.~R.~Hiller and G.~McCartor,
Phys. Rev. D \textbf{58}, 025005 (1998)
doi:10.1103/PhysRevD.58.025005
[arXiv:hep-th/9802120 [hep-th]].

\bibitem{Karmanov:2008br}
V.~A.~Karmanov, J.~F.~Mathiot and A.~V.~Smirnov,
Phys. Rev. D \textbf{77}, 085028 (2008)
doi:10.1103/PhysRevD.77.085028
[arXiv:0801.4507 [hep-th]].

\bibitem{Karmanov:2016yzu}
V.~A.~Karmanov, Y.~Li, A.~V.~Smirnov and J.~P.~Vary,
Phys. Rev. D \textbf{94}, no.9, 096008 (2016)
doi:10.1103/PhysRevD.94.096008
[arXiv:1610.03559 [hep-th]].

\bibitem{Karmanov:2012aj}
V.~A.~Karmanov, J.~F.~Mathiot and A.~V.~Smirnov,
Phys. Rev. D \textbf{86}, 085006 (2012)
doi:10.1103/PhysRevD.86.085006
[arXiv:1204.3257 [hep-th]].

\bibitem{Kadyshevsky:1967rs}
V.~G.~Kadyshevsky,
Nucl. Phys. B \textbf{6}, 125-148 (1968)
doi:10.1016/0550-3213(68)90274-5

\bibitem{Atakishiev:1976sa}
N.~M.~Atakishiev, R.~M.~Mir-Kasimov and S.~M.~Nagiev,
Teor. Mat. Fiz. \textbf{32}, 34-43 (1977)
doi:10.1007/BF01041430

\bibitem{Karmanov:1976iv}
V.~A.~Karmanov,
Zh. Eksp. Teor. Fiz. \textbf{71}, 399-416 (1976)
ITEP-54-1976.

\bibitem{Lehmann:1954rq}
H.~Lehmann, K.~Symanzik and W.~Zimmermann,
Nuovo Cim. \textbf{1}, 205-225 (1955)
doi:10.1007/BF02731765

\bibitem{Lehmann:1957zz}
H.~Lehmann, K.~Symanzik and W.~Zimmermann,
Nuovo Cim. \textbf{6}, 319-333 (1957)
doi:10.1007/BF02832508

\bibitem{Karmanov:1991fv}
V.~A.~Karmanov and A.~V.~Smirnov,
Nucl. Phys. A \textbf{546}, 691-717 (1992)
doi:10.1016/0375-9474(92)90004-4

\bibitem{Karmanov:1994ck}
V.~A.~Karmanov and A.~V.~Smirnov,
Nucl. Phys. A \textbf{575}, 520-548 (1994)
doi:10.1016/0375-9474(94)90374-3

\bibitem{Karmanov:1996un}
V.~A.~Karmanov and J.~F.~Mathiot,
Nucl. Phys. A \textbf{602}, 388-404 (1996)
doi:10.1016/0375-9474(96)00101-7

\bibitem{Carbonell:1999pt}
J.~Carbonell and V.~A.~Karmanov,
Eur. Phys. J. A \textbf{6}, 9-19 (1999)
doi:10.1007/s100500050311
[arXiv:nucl-th/9902053 [nucl-th]].

\bibitem{Brodsky:2003pw}
S.~J.~Brodsky, J.~R.~Hiller, D.~S.~Hwang and V.~A.~Karmanov,
Phys. Rev. D \textbf{69}, 076001 (2004)
doi:10.1103/PhysRevD.69.076001
[arXiv:hep-ph/0311218 [hep-ph]].

\bibitem{Choi:2021xld}
Y.~Choi, H.~M.~Choi, C.~R.~Ji and Y.~Oh,
Phys. Rev. D \textbf{103}, no.7, 076002 (2021)
doi:10.1103/PhysRevD.103.076002
[arXiv:2101.03656 [hep-ph]].

\bibitem{Smirnov:2010}
A.V. Smirnov, private communication. 

\bibitem{Heinzl:2018xnv}
T.~Heinzl, A.~Ilderton and D.~Seipt,
Phys. Rev. D \textbf{98}, no.1, 016002 (2018)
doi:10.1103/PhysRevD.98.016002
[arXiv:1802.05933 [hep-th]].

\bibitem{Chabysheva:2015vga}
S.~S.~Chabysheva and J.~R.~Hiller,
[arXiv:1506.05429 [hep-th]].

\bibitem{Hiller:2015bic}
J.~R.~Hiller,
Few Body Syst. \textbf{57}, no.7, 557-563 (2016)
doi:10.1007/s00601-016-1098-9
[arXiv:1512.08771 [hep-th]].





\bibitem{Chabysheva:2011ed}
S.~S.~Chabysheva and J.~R.~Hiller,
Phys. Lett. B \textbf{711}, 417-422 (2012)
doi:10.1016/j.physletb.2012.04.032
[arXiv:1103.0037 [hep-ph]].


\bibitem{Glazek:1993rc}
S.~D.~Glazek and K.~G.~Wilson,
Phys. Rev. D \textbf{48}, 5863-5872 (1993)
doi:10.1103/PhysRevD.48.5863


\bibitem{Glazek:1994qc}
S.~D.~Glazek and K.~G.~Wilson,
Phys. Rev. D \textbf{49}, 4214-4218 (1994)
doi:10.1103/PhysRevD.49.4214

\bibitem{Wilson:1994fk}
K.~G.~Wilson, T.~S.~Walhout, A.~Harindranath, W.~M.~Zhang, R.~J.~Perry and S.~D.~Glazek,
Phys. Rev. D \textbf{49}, 6720-6766 (1994)
doi:10.1103/PhysRevD.49.6720
[arXiv:hep-th/9401153 [hep-th]].


\end{thebibliography}
\end{document}